\newif\ifelsevier
\elsevierfalse

\ifelsevier
\documentclass{elsart}
\else
\documentclass[11pt,letterpaper]{article}
\fi
\usepackage{amsmath}
\usepackage{mathtools}
\usepackage{amsfonts}
\usepackage{amssymb}
\usepackage{graphicx}
\usepackage{color}
\ifelsevier
\newenvironment{proof}[1][Proof.]{\begin{pf*}{#1}}{\qed\end{pf*}}
\else
\usepackage{amsthm}
\fi
\usepackage{hyperref}

\newtheorem{theorem}{Theorem}
\newtheorem{lemma}[theorem]{Lemma}
\newtheorem{conjecture}{Conjecture}

\newtheorem{definition}{Definition}

\newcommand{\eps}{\epsilon}

\newcommand{\card}[1]{\lvert#1\rvert}
\newcommand{\suchthat}{\mathrel{:}}
\newcommand{\giventhat}{\mid}

\newcommand{\RR}{\mathbb{R}}
\newcommand{\R}{\RR}

\newcommand{\norm}[1]{{\lVert#1\rVert}}

\newcommand{\abs}[1]{\lvert#1\rvert}
\newcommand{\vol}{\operatorname{vol}}

\newcommand{\inner}[2]{#1\cdot #2}

\newcommand{\expectation}{\operatorname{\mathbb{E}}}
\newcommand{\E}{\expectation}
\newcommand{\var}{\operatorname{var}}
\newcommand{\cov}[2]{\operatorname{cov}({#1}, {#2})}
\newcommand{\grad}{\nabla}

\providecommand{\e}{\expectation}
\renewcommand{\e}{\expectation}
\newcommand{\norms}[1]{{\norm{#1}}^2}
\newcommand{\inorm}[1]{\lVert{#1}\rVert_\infty}
\newcommand{\sign}{\operatorname{sign}}
\newcommand{\proj}{\pi}
\newcommand{\disp}[2]{\operatorname{disp}_{#1}\left(#2\right)}

\newcommand{\poly}{\operatorname{poly}}
\newcommand{\mycase}[1]{\medskip\noindent\emph{Case {#1}:} }
\newcommand{\killtext}[1]{}
\newcommand{\email}[1]{\protect\href{mailto:#1}{\nolinkurl{#1}}}

\begin{document}
\ifelsevier
    \begin{frontmatter}
\fi
    \title{Dispersion of Mass and the Complexity of Randomized Geometric Algorithms}
\ifelsevier
    \thanks[lr]{Supported by NSF ITR-0312354. At MIT when this research was performed.}
    \thanks[sv]{Supported in part by a Guggenheim fellowship and by NSF CCF-0721503. At MIT when this research was performed.}
    \author{Luis Rademacher\corauthref{cor}\thanksref{lr}}
    \address{Georgia Tech, Atlanta, GA 30332}
    %\address{MIT, Cambridge, MA 02139}
    \ead{lrademac@cc.gatech.edu}
    \author{Santosh Vempala\thanksref{sv}}
    \address{Georgia Tech, Atlanta, GA 30332}
    \ead{vempala@cc.gatech.edu} \corauth[cor]{Corresponding author.}
\else
    \author{Luis Rademacher\thanks{Supported by NSF ITR-0312354. At MIT when this research was performed.}\\
    Georgia Tech\\
    \email{lrademac@cc.gatech.edu} \and
    Santosh Vempala\thanks{Supported in part by a Guggenheim fellowship and by NSF CCF-0721503. At MIT when this research was performed.}\\
    Georgia Tech\\
    \email{vempala@cc.gatech.edu}}
    \date{}
    \maketitle
\fi
\begin{abstract}
How much can randomness help computation? Motivated by this general
question and by volume computation, one of the few instances where
randomness provably helps, we analyze a notion of dispersion and
connect it to asymptotic convex geometry. We obtain a nearly
quadratic lower bound on the complexity of randomized volume
algorithms for convex bodies in $\R^n$ (the current best algorithm
has complexity roughly $n^4$, conjectured to be $n^3$). Our main
tools, dispersion of random determinants and dispersion of the
length of a random point from a convex body, are of independent
interest and applicable more generally; in particular, the latter is
closely related to the variance hypothesis from convex geometry.
This geometric dispersion also leads to lower bounds for matrix
problems and property testing.
\end{abstract}

\ifelsevier
\end{frontmatter}
\fi
%\thispagestyle{empty}
%\newpage \setcounter{page}{1}

\section{Introduction}

Among the most intriguing questions raised by complexity theory is the
following: how much can the use of randomness affect the computational
complexity of algorithmic problems? At the present time, there are many problems
for which randomized algorithms are simpler or faster than known deterministic
algorithms but only a few known instances where randomness provably helps.

One problem for which randomness makes a dramatic difference is
estimating the volume of a convex body in $\R^n$. The convex body
can be accessed as follows: for any point $x \in \R^n$, we can
determine whether $x$ is in the body or not (a {\em membership}
oracle). The complexity of an algorithm is measured by the number of
such queries. The work of Elekes \cite{El} and B\'{a}r\'{a}ny and
F\"{u}redi \cite{BF} showed that any deterministic polynomial-time
algorithm cannot estimate the volume to within an exponential (in
$n$) factor. We quote their theorem below.

\begin{theorem}[\cite{BF}]
For every deterministic algorithm that uses at most $n^a$ membership queries
and given a convex body $K$ with $B_n \subseteq K \subseteq nB_n$ outputs
two numbers $A,B$ such that $A \leq \vol(K) \leq B$,
there exists a body $K'$
for which the ratio $B/A$ is at least
\[
\left(\frac{cn}{a\log n}\right)^n
\]
where $c$ is an absolute constant.
\end{theorem}

In striking contrast, the celebrated paper of Dyer, Frieze and
Kannan \cite{DFK} gave a polynomial-time randomized algorithm to
estimate the volume to arbitrary accuracy (the dependence on $n$ was
about $n^{23}$). This result has been much improved and generalized
in subsequent work ($n^{16}$, \cite{LS90}; $n^{10}$,
\cite{Lov90,AK}; $n^{8}$, \cite{DF}; $n^{7}$, \cite{LS93}; $n^{5}$,
\cite{KLS97}; $n^{4}$, \cite{LV03}); the current fastest algorithm
has complexity that grows as roughly $O(n^4/\eps^2)$ to estimate the
volume to within relative error $1+\eps$ with high probability (for
recent surveys, see \cite{Sim,Vem05}). Each improvement in the
complexity has come with fundamental insights and lead to new
isoperimetric inequalities, techniques for analyzing convergence of
Markov chains, algorithmic tools for rounding and sampling
logconcave functions, etc..

These developments lead to the question: what is the best possible complexity
of any randomized volume algorithm? A lower bound of $\Omega(n)$ is straightforward.
Here we prove a nearly quadratic lower bound: there is a constant $c > 0$
such that any randomized algorithm
that approximates the volume to within a $(1+c)$ factor
needs $\Omega(n^2/\log n)$
queries. The formal statement appears in Theorem \ref{thm:volumeLowerBound}.

For the more restricted class of randomized nonadaptive algorithms
(also called ``oblivious''), an exponential lower bound is
straightforward (Section \ref{nonadaptive}). Thus, the use of
full-fledged adaptive randomization is crucial in efficient volume
estimation, but cannot improve the complexity below $n^2/\log n$.

In fact, the quadratic lower bound holds for a restricted class of
convex bodies, namely parallelopipeds. A parallelopiped in $\R^n$
centered at the origin can be compactly represented using a matrix
as $\{x : \norm{Ax}_\infty \le 1\}$, where $A$ is an $n\times n$
nonsingular matrix;  the volume is simply $2^n \abs{\det(A)}^{-1}$.
One way to interpret the lower bound theorem is that in order to
estimate $\abs{\det(A)}$ one needs almost as many bits of
information as the number of entries of the matrix. The main
ingredient of the proof is a dispersion lemma which shows that the
determinant of a random matrix remains dispersed even after
conditioning the distribution considerably. We discuss other
consequences of the lemma in Section \ref{DISCUSS}.

Our lower bound is nearly the best possible for this restricted
class of convex bodies.  Using $O(n^2\log n)$ queries, we can find a
close approximation to the entire matrix $A$ and therefore any
reasonable function of its entries. This naturally raises the
question of what other parameters require a quadratic number of
queries. We prove that estimating the product of the lengths of the
rows of an unknown matrix $A$ to within a factor of about $(1+1/\log
n)$ also requires $\Omega(n^2/\log n)$ queries. The simplest version
of this problem is the following: given a membership oracle for any
unknown halfspace $a\cdot x \le 1$, estimate $\norm{a}$, the
Euclidean length of the normal vector $a$ (alternatively, estimate
the distance of the hyperplane from the origin). This problem can be
solved deterministically using $O(n\log n)$ oracle queries. We prove
that any randomized algorithm that estimates $\norm{a}$ to within an
additive error of about $1/\sqrt{\log n}$ requires $\Omega(n)$
oracle queries.

Related earlier work includes \cite{BLY,DL}, showing lower bounds
for linear decision trees (i.e., every node of the tree tests
whether an affine function of the input is nonnegative). \cite{BLY}
considers the problem of deciding whether given $n$ real numbers,
some $k$ of them are equal, and they prove that it has complexity
$\Theta(n \log (n/k))$. \cite{DL} proves that the $n$-dimensional
knapsack problem has complexity at least $n^2/2$.

For these problems (length, product of lengths), the main tool in
the analysis is a geometric dispersion lemma that is of independent
interest in asymptotic convex geometry. Before stating the lemma, we
give some background and motivation. There is an elegant body of
work that studies the distribution of a random point $X$ from a
convex body $K$ \cite{Ba,BK,Bou,MP}. A convex body $K$ is said to be
in {\em isotropic} position if $\vol(K)=1$ and for a random point
$X$ we have
\[
\E(X) = 0, \quad \mbox{ and } \quad \E(XX^T) = \alpha I \mbox{ for some }\alpha > 0
.
\]
We note that there is a slightly different definition of isotropy
(more convenient for algorithmic purposes) which does not restrict
$\vol(K)$ and replaces the second condition above by $\E(XX^T)=I$.
Any convex body can be put in isotropic position by an affine
transformation. A famous conjecture ({\em isotropic constant}) says
that $\alpha$ is bounded by a universal constant for every convex
body. It follows that $\E(\norm{X}^2) = O(n)$. Motivated by the
analysis of random walks, Lov\'{a}sz and Vempala made the following
conjecture (under either definition). If true, then some natural
random walks are significantly faster for isotropic convex bodies.
\begin{conjecture}\label{conj:variance}
For a random point $X$ from an isotropic convex body,
\[
\var(\norms{X}) = O(n).
\]
\end{conjecture}
%As we discuss in Section \ref{sec:VarPoly}, this conjecture is implied by
%the other famous conjectures in convex geometry.
The upper bound of $O(n)$ is
achieved, for example, by the isotropic cube. The isotropic ball, on the other
hand, has the smallest possible value, $\var(\norms{X}) = O(1)$. The
variance lower bound we prove in this paper (Theorem \ref{VarPoly})
directly implies the following: for an isotropic convex polytope $P$
in $\R^n$ with at most $\poly(n)$ facets,
\[
\var(\norms{X}) = \Omega\left(\frac{n}{\log n}\right).
\]
Thus, the conjecture is nearly
tight for not just the cube, but {\em any} isotropic polytope with a
small number of facets. Intuitively, our lower bound shows that the
length of a random point from such a polytope is {\em not
concentrated} as long as the volume is reasonably large. Roughly speaking,
this says that in order to determine the length, one would have to
localize the entire vector in a small region.

Returning to the analysis of algorithms, one can
view the output of a randomized algorithm as a distribution. Proving a lower
bound on the complexity is then equivalent to showing that the
output distribution after some number of steps
is {\em dispersed}.
To this end, we define a simple parameter of a distribution:
\begin{definition}
Let $\mu$ be a probability measure on $\RR$. For any $0 < p < 1$, the
{\em $p$-dispersion} of $\mu$ is
\[
\disp{\mu}{p} = \inf \{|a-b| \suchthat a,b \in \RR, \mu([a,b]) \geq
1-p\}.
\]
\end{definition}
Thus, for any possible output $z$, and a random point $X$, with
probability at least $p$,
$|X-z| \ge \disp{\mu}{p}/2$.
We prove some useful properties about this parameter in Section \ref{PRELIM}.

\section{Results}

\subsection{Complexity lower bounds}
We begin with our lower bound for randomized
volume algorithms. Besides the dimension $n$,
the complexity also depends on the ``roundness'' of the input body.
This is the ratio $R/r$ where $rB_n \subseteq K \subseteq RB_n$.
To avoid another parameter in our
results, we ensure that $R/r$ is bounded by a polynomial in $n$.
\begin{theorem}[volume]\label{thm:volumeLowerBound}
Let $K$ be a convex body given by a membership oracle such that $B_n
\subseteq K \subseteq O(n^{8})B_n$. Then there exists a constant
$c>0$ such that any randomized algorithm that outputs a number $V$
such that $(1-c)\vol(K) \le V \le (1+c)\vol(K)$ holds with
probability at least $1-1/n$ has complexity $\Omega(n^2/\log n)$.
\end{theorem}
\noindent We note that the lower bound can be easily extended to any
algorithm with success probability $p > 1/2$ with a small overhead
\cite{JVV}. The theorem actually holds for parallelopipeds with the
same roundness condition, i.e., convex bodies specified by an $n
\times n$ real matrix $A$ as $\{ x \in \RR^n \suchthat \forall\ 1
\le i \le n \quad \abs{A_i \cdot x} \leq 1\}$ where $A_i$ denotes
the $i$'th row of $A$. In this case, the volume of $K$ is simply
$2^n \abs{\det(A)}^{-1}$. We restate the theorem for this case.
\begin{theorem}[determinant]\label{thm:determinantLowerBound}
Let $A$ be an matrix with entries in $[-1,1]$ and smallest singular
value at least $2^{-12}n^{-7}$ that can be accessed by the following
oracle: for any $x$, the oracle determines whether $\norm{Ax}_\infty
\le 1$ is true or false. Then there exists a constant $c>0$ such
that any randomized algorithm that outputs a number $V$ such that
\[
(1-c)\abs{\det(A)} \le V \le (1+c)\abs{\det(A)}
\]
holds with probability at least $1-1/n$, has complexity
$\Omega(n^2/\log n)$.
\end{theorem}
A slightly weaker lower bound holds for estimating
the product of the lengths of the rows of $A$. The proof is in
Section \ref{sec:product}.
\begin{theorem}[product]\label{PRODUCT}
Let $A$ be an unknown matrix that can be accessed by the following oracle:
for any $x$, the oracle determines whether $||Ax||_\infty \le 1$ is true
or false. Then there exists a constant
$c>0$ such that any randomized algorithm that outputs a number $L$
such that
\[
\left(1-\frac{c}{\log n}\right)\prod_{i=1}^n\norm{A_i} \le L \le
\left(1+\frac{c}{\log n}\right)\prod_{i=1}^n\norm{A_i}
\]
with probability at least $1-1/n$ has complexity $\Omega(n^2/\log n)$.
\end{theorem}
When $A$ has only a single row, we get a stronger bound. In this case,
the oracle is simply a membership oracle for a halfspace.
\begin{theorem}[length]\label{thm:lengthLowerBound}
Let $a$ be a vector in $[-1,1]^n$ with $\norm{a} \ge
\sqrt{n}-4\sqrt{\log n}$ and $a \cdot x \le 1$ be the corresponding
halfspace in $\R^n$ given by a membership oracle. Then there exists
a constant $c > 0$ such that any randomized algorithm that outputs a
number $l$ such that
\[
\norm{a}-\frac{c}{\sqrt{\log n}} \leq l \leq \norm{a}+\frac{c}{\sqrt{\log n}}
\]
with probability at least $1 - 1/n$ has complexity at least $n-1$.
\end{theorem}
The restrictions on the input in all the above theorems (``roundness'') only
make them stronger. For example, the bound on the length of $a$ above
implies that it only varies in
an interval of length $4\sqrt{\log n}$. To pin it down in an
interval of length $c/\sqrt{\log n}$ (which is $O(\log\log n)$ bits
of information) takes $\Omega(n)$
queries. This result is in the spirit of hardcore predicates
\cite{GL89}.

It is worth noting that a very simple algorithm can approximate the
length as in the theorem with probability at least $3/4$ and $O(n
\log^2 n)$ queries: the projection of $a$ onto a given vector $b$
can be computed up to an additive error of $1/\poly(n)$ in $O(\log
n)$ queries (binary search along the line spanned by $b$). If $b$ is
random in $S_{n-1}$, then $\E ((\inner{a}{b})^2) = \norms{a}/n$. A
Chernoff-type bound gives that the average of $O(n \log n)$ random
projections allows the algorithm to localize $\norm{a}$ in an
interval of length $O(1/\sqrt{\log n})$ with probability at least
$3/4$.

\subsection{Variance of polytopes}
The next theorem states that the length of a random
point from a polytope with few facets
has large variance. This is a key tool in our
lower bounds. It also has a
close connection to the variance hypothesis (which
conjectures an upper bound for all isotropic convex bodies), suggesting
that polytopes might be the limiting case of that conjecture.
\begin{theorem}\label{VarPoly}\label{thm:variancePolytope}
Let $P \subseteq \RR^n$ be a polytope
with at most $n^k$ facets and contained in the ball of radius
$n^q$. For a random point $X$ in $P$,
\[
\var \norms{X} \geq
\vol(P)^{\frac{4}{n}+\frac{3c}{n\log n}}
e^{-c(k+3q)}\frac{n}{\log n}
\]
where $c$ is a universal constant.
\end{theorem}
Thus, for a polytope of volume at least $1$ contained in a ball of
radius at most $\poly(n)$, with at most $\poly(n)$ facets, we have
$\var \norms{X} = \Omega(n/\log n)$. In particular this holds for
any isotropic polytope with at most $\poly(n)$ facets. The proof
of Theorem \ref{thm:variancePolytope} is given in Section \ref{sec:VarPoly}.

\subsection{Dispersion of the determinant}
In our proof of the volume lower bound, we begin with a distribution on matrices for which
the determinant is dispersed. The main goal of the proof is to show that even after considerable conditioning, the determinant is still dispersed. The next definition will be
useful in describing the structure of the distribution and how it changes
with conditioning.

\begin{definition}\label{def:productSet}
Let ${\cal M}$ be a set of $n \times n$ matrices. We say that ${\cal
M}$ is a {\em product set along rows} if there exist sets ${\cal
M}_i \subseteq \R^n$, $1 \le i \le n$,
\[
{\cal M} = \{ M : \forall 1 \le i \le n, M_i \in {\cal M}_i\}.
\]
\end{definition}
Let $B_n$ denote the $n$-dimensional Euclidean unit ball centered at
the origin.
\begin{lemma}\label{thm:determinantDispersion}
There exists a constant $c>0$ such that for any partition $\{
\mathcal{A}^j\}_{j \in N}$ of $(\sqrt{n}B_n)^n$ into $\card{N} \leq
2^{n^2-2}$ parts where each part is a product set along rows, there
exists a subset $N' \subseteq N$ such that
\begin{enumerate}
\item[a.] $\vol(\bigcup_{j \in N'} {\cal A}^j) \ge \frac{1}{2}
\vol\bigl((\sqrt{n}B_n)^n\bigr)$ and

\item[b.] for any $u> 0$ and a random point $X$ from ${\cal A}^j$ for any
$j \in N'$, we have
\[
\Pr\bigl( \abs{\det{X}} \notin [u,u(1+c)] \bigr) \geq \frac{1}{2^{7}
n^6}.
\]
\end{enumerate}
\end{lemma}
%\comment{\[\mdisp{X \in P}{1-\frac{1}{2^{16} n^4}} \geq c\]}

\section{Preliminaries}\label{PRELIM}

Throughout the paper, we assume that $n > 12$ to avoid trivial
complications.

We define $\pi_V(u)$ to be the projection of a vector $u$ to a
subspace $V$. Given a matrix $R$, let $R_i$ denote the $i$'th row of
$R$, and let $\hat R$ be the matrix having the rows of $R$
normalized to be unit vectors. Let
%$\tilde R_i = \proj_{(R_1, \dotsc, R_{i-1})^\perp}(R_i)$.
$\tilde R_i$ be the projection of $R_i$ to the subspace orthogonal
to $R_1, \dotsc, R_{i-1}$. For any row $R_i$ of matrix $R$, let
$R_{-i}$ denote (the span of) all rows except $R_i$. So
$\proj_{R_{-i}^\perp}(R_i)$ is the projection of $R_i$ orthogonal to
the subspace spanned by all the other rows of $R$.

The volume of the Euclidean unit ball is given by
$\pi^{n/2}/\Gamma(n/2+1)$, and its surface area is
$2\pi^{n/2}/\Gamma(n/2)$.

\subsection{Dispersion}
%\begin{definition}
%Let $\mu$ be a probability measure on $[0,+\infty)$. For any $0 < p
%< 1$, the \emph{multiplicative $p$-dispersion} of $\mu$ is
%\[
%\mdisp{\mu}{p} = \inf \{\epsilon \geq 0 \suchthat (\exists a \geq 0)
%\mu([a,a(1+\epsilon)]) \geq p\}
%\]
%\end{definition}
%Because every real random variable induces a probability measure on
%$\RR$, we can also talk about the dispersion of a real random
%variable.
%\[
%\disp{X}{p} = \log (1+\mdisp{e^X}{p}) ???
%\]
We begin with
two simple cases in which large variance implies large dispersion.
\begin{lemma}\label{lem:varImpliesDisp}
Let $X$ be a real random variable with finite variance $\sigma^2$.
\begin{itemize}
\item[a.] If the support of $X$ is contained in an interval of
    length $M$ then
    \[
        \disp{X}{\frac{3\sigma^2}{4M^2}} \geq \sigma.
    \]

\item[b.] If $X$ has a logconcave density then
$\disp{X}{p} \geq (1-p) \sigma$.
\end{itemize}
\end{lemma}

\begin{proof}
Let $a,b \in \RR$ be such that $b-a < \sigma$. Let $\alpha = \Pr(X
\notin [a,b])$. Then
\[
\var X \leq (1-\alpha)\left(\frac{b-a}{2}\right)^2 + \alpha M^2.
\]
This implies
\[
\alpha > \frac{3 \sigma^2}{4M^2}.
\]

For the second part, Lemma 5.5(a) from \cite{lvlogconcave} implies
that a logconcave density with variance $\sigma^2$ is never greater
than $1/\sigma$. This implies that if $a,b \in \RR$ are such that
$\Pr(X \in [a,b]) \geq p$ then we must have $b-a \geq p \sigma$.
\end{proof}

%\begin{lemma}
%Let $X$, $Y$ be independent real random variables with logconcave
%densities. Then
%\[
%\disp{X+Y}{p} \geq p \sqrt{\var(X) + \var(Y)}
%\]
%\end{lemma}
%\begin{proof}
%The hypotheses imply that $\var(X + Y) = \var(X) + \var(Y)$. Further,
%the density of $X+Y$ is the convolution of the densities of $X$ and $Y$; since
%the convolution of logconcave functions is logconcave,
%$X+Y$ also has a logconcave density. Lemma
%\ref{lem:varImpliesDisp}(b) gives the desired conclusion.
%\end{proof}

\begin{lemma}
Let $X,Y$ be real-valued random variables and $Z$ be a random
variable that is generated by setting it equal to $X$ with
probability $\alpha$ and equal to $Y$ with probability $1-\alpha$.
Then,
\[
\disp{Z}{\alpha p} \ge \disp{X}{p}.
\]
\end{lemma}

%Connection with failure of deterministic algorithms against a
%distribution.

%Connection with existence of 2 disjoint intervals with some mass and
%far apart.

\begin{lemma}\label{UNIFORMPART}
Let $f:[0,M]\rightarrow\R_+$ be a density function with mean $\mu$
and variance $\sigma^2$. Suppose the distribution function of $f$ is
logconcave. Then $f$ can be decomposed into a convex combination of
densities $g$ and $h$, i.e., $f(x)=\alpha g(x) + (1-\alpha)h(x)$,
where $g$ is uniform over an interval $[a,b]$, with $a \ge \mu$ and
$\alpha (a-b)^2 = \Omega\bigl(\sigma^2/\log (M/\sigma)\bigr)$.
%where $g$ is uniform over an interval $[a,b]$, with $a \ge \mu$,
%$\alpha (a-b)^2 = \Omega\bigl(\sigma^2/\log (M/\sigma)\bigr)$ and
%$\alpha =\Omega\bigl(\sigma^2/M^2\log (M/\sigma)\bigr)$.
\end{lemma}
This lemma is proved in Section \ref{sec:product}.

\subsection{Yao's lemma}
We will need the following version of Yao's lemma.
Informally, the probability of failure of a
randomized algorithm $\nu$ on the worst input is at least the
probability of failure of the best deterministic algorithm against
some distribution $\mu$.
\begin{lemma}\label{lem:yao2}
Let $\mu$ be a probability measure on inputs $I$ (a ``distribution
on inputs'') and let $\nu$ be a probability measure on deterministic
algorithms $A$ (a ``randomized algorithm''). Then
\begin{multline*}
\inf_{a \in A} \Pr (\text{algorithm $a$ fails on measure $\mu$}) \\
\leq \sup_{i \in I} \Pr(\text{randomized algorithm $\nu$ fails on
input $i$}).
\end{multline*}
\end{lemma}

Let $I$ be a set (a subset of the inputs of a computational problem,
for example the set of all well-rounded convex bodies in $\RR^n$ for
some $n$). Let $O$ be another set (the set of possible outputs of a
computational problem, for example, real numbers that are an
approximation to the volume of a convex body). Let $A$ be a set of
functions from $I$ to $O$ (these functions represent deterministic
algorithms that take elements in $I$ as inputs and have outputs in
$O$). Let $C : I \times A \to \RR$ (for $a \in A$ and $i \in I$,
$C(i,a)$ is a measure of the badness of the algorithm $a$ on input
$i$, such as the indicator of $a$ giving a wrong answer on $i$).
\begin{lemma}
\label{lem:yao}
Let $\mu$ and $\nu$ be probability measures
over $I$ and $A$, respectively. Let $C : I \times A \to \RR$ be
integrable with respect to $\mu \times \nu$. Then
\begin{align*}
\inf_{a \in A} \expectation_{\mu(i)} C(i,a) \leq \sup_{i \in I}
\expectation_{\nu(a)} C(i,a)
\end{align*}
\end{lemma}
%(one could prove an alternative version that replaces the
%integrability of $C$ by measurability and non-negativity)
\begin{proof}
By means of Fubini's theorem and the integrability assumption we
have
\[
\e_{\nu(a)} \e_{\mu(i)} C(i,a) = \e_{\mu(i)} \e_{\nu(a)} C(i,a).
\]
Also
\[
\e_{\nu(a)} \e_{\mu(i)} C(i,a) \geq \inf_{a \in A} \e_{\mu(i)}
C(i,a)
\]
and
\[
\e_{\mu(i)} \e_{\nu(a)} C(i,a) \leq \sup_{i \in I} \e_{\nu(a)}
C(i,a).
\]
\end{proof}

\begin{proof}[Proof (of Lemma \ref{lem:yao2})]
Let $C : I \times A \to \RR$, where for $i \in I$, $a \in A$ we have
\[
C(i,a) = \begin{cases}
    1 & \text{if $a$ fails on $i$} \\
    0 & \text{otherwise.}
\end{cases}
\]
Then the consequence of Lemma \ref{lem:yao} for this $C$ is
precisely what we want to prove.
\end{proof}

\subsection{The query model and decision trees}
We have already discussed the standard query model (let us call it
$Q$): A membership oracle for a convex body $K$ takes any $q \in
\RR^n$ and outputs YES if $q \in K$ and NO otherwise. When $K$ is a
parallelopiped specified by a matrix $A$, the oracle outputs YES if
$\inorm{A q} \leq 1$ and NO otherwise.

It is useful to view the computation of a deterministic algorithm as
a decision tree representing the sequence of queries: the nodes
(except the leaves) represent queries, the root is the first query
made by the algorithm and there is one query subtree per answer. The
leaves do not represent queries but instead the answers to the last
query along every path. Any leaf $l$ has a set $P_l$ of inputs that
are consistent with the corresponding path of queries and answers on
the tree. Thus the set of inputs is partitioned by the leaves.

To prove our main lower bound results for parallelopipeds, it will
be convenient to consider a modified query model $Q'$ that can
output more information: Given $q \in \RR^n$, the modified oracle
outputs YES as before if $\inorm{Aq} \leq 1$; otherwise it outputs a
pair $(i,s)$ where $i$ is the ``least index among violated
constraints'', $i = \min \{ j \suchthat \abs{A_j q} > 1\}$, and $s
\in \{-1,1\}$ is the ``side'', $s = \sign (A_i q)$. An answer from
$Q'$ gives at least as much information as the respective answer
from $Q$, and this implies that a lower bound for algorithms with
access to $Q'$ is also a lower bound for algorithms with access to
$Q$. The modified oracle $Q'$ has the following useful property (see
Definition \ref{def:productSet}):
\begin{lemma}\label{LEM:TREE}
If the set of inputs is a product set along rows, then the leaves of
a decision tree in the modified query model $Q'$ induce a partition
of the input set where each part is itself a product set along rows.
\end{lemma}

\begin{proof}
We start with ${\cal M}$, a product set along rows with components
${\cal M}_i$. Let us observe how this set is partitioned as we go
down a decision tree. A YES answer imposes two additional
constraints of the form $-1 \le q\cdot x \le 1$ on every set ${\cal
M}_i$. For a NO answer with response $(i,s)$, we get two constraints
for all ${\cal M}_j$, $1 \le j < i$, one constraint for the $i$'th
set and no new constraints for the remaining sets. Given this
information, a particular setting of any row (or subset of rows)
gives no additional information about the other rows. Thus, the set
of possible matrices at each child of the current query is a product
set along rows. The lemma follows by applying this argument
recursively.
\end{proof}

Apart from the product property given by the previous lemma, if one
assumes additionally that the set of inputs is convex, then in the
query model $Q'$ each part of the partition is a convex set. This
property is used in the proof of the product lower bound (Theorem
\ref{PRODUCT}), but is not used in the volume lower bound (Theorem
\ref{thm:volumeLowerBound}). Thus, for the volume lower bound one
could use an oracle like $Q'$ that outputs the index $i$ but not the
sign $s$, and the product property would be preserved (Lemma
\ref{LEM:TREE}) but not the convexity.

\subsection{Distributions and concentration properties}
We use two distributions on $n \times n$ matrices called $D$ and $D'$
for the lower bounds in this paper. A random matrix from $D$ is obtained
by selecting each row independently and uniformly from the ball of radius
$\sqrt{n}$. A random matrix from $D'$ is obtained by selecting each entry of
the matrix independently and uniformly from the interval $[-1,1]$.
In the analysis, we will also encounter random matrices where each
entry is selected independently from $N(0,1)$. We use the following
property.
\begin{lemma}\label{MINSINGULAR}
Let $\sigma$ be
the minimum singular value of an $n \times n$ matrix $G$
with independent entries from $N(0,1)$. For any $t > 0$,
\[
\Pr\left(\sigma\sqrt{n} \le t\right) \le t.
\]
\end{lemma}
\begin{proof}
To bound $\sigma$, we will consider the formula for the density of
$\lambda = \sigma^2$ given in \cite[Theorem 3.1]{Edel}:
\[
f(\lambda) = \frac{n}{2^{n-1/2}} \frac{\Gamma(n)}{\Gamma(n/2)}
\lambda^{-1/2} e^{-\lambda n/2} U\left(\frac{n-1}{2}, -\frac{1}{2},
\frac{\lambda}{2}\right)
\]
where $U$ is the Tricomi function, which satisfies for all $\lambda
\geq 0$:
\begin{itemize}
\item $U(\frac{n-1}{2}, -\frac{1}{2},
0)  = \Gamma(3/2)/\Gamma((n+2)/2)$,

\item $U(\frac{n-1}{2}, -\frac{1}{2},
\lambda) \geq 0$

\item $\frac{d}{d\lambda} U(\frac{n-1}{2}, -\frac{1}{2},
\lambda)\leq 0$
\end{itemize}
(The first two properties are from \cite[Theorem 3.1]{Edel}, the
third from \cite[13.1.3 and 13.4.21]{AandS}.)

We will now prove that for any $n$ the density function of $t =
\sqrt{n \lambda}$ is at most $1$. To see this, the density of $t$ is
given by
\begin{align*}
g(t) &= f\left(\frac{t^2}{n}\right) \frac{2t}{n} = 2 f(\lambda)
\sqrt{\frac{\lambda}{n}}
    = \frac{\sqrt{n}}{2^{n-3/2}} \frac{\Gamma(n)}{\Gamma(n/2)}
 e^{-\lambda n/2} U\left(\frac{n-1}{2}, -\frac{1}{2},
\frac{\lambda}{2}\right).
\end{align*}
Now,
\begin{multline*}
\frac{d}{dt} g(t) = \frac{\sqrt{n}}{2^{n-3/2}}
\frac{\Gamma(n)}{\Gamma(n/2)} \times \\
\times \left[-\frac{n}{2} e^{-\lambda n/2} U\left(\frac{n-1}{2},
-\frac{1}{2}, \frac{\lambda}{2}\right) + e^{-\lambda n/2}
\frac{d}{d\lambda}U\left(\frac{n-1}{2}, -\frac{1}{2},
\frac{\lambda}{2}\right)\right] \frac{2t}{n} \leq 0.
\end{multline*}
Thus, the maximum of $g$ is at $t=0$, and
\begin{align*}
g(0) &= \frac{\sqrt{n}}{2^{n-3/2}} \frac{\Gamma(n)}{\Gamma(n/2)}
 \frac{\Gamma(3/2)}{\Gamma(\frac{n+2}{2})} \leq 1.
\end{align*}
It follows that $\Pr(\sigma \sqrt{n} \leq \alpha) \leq \alpha$.
\end{proof}

\begin{lemma}\label{lem:gaussianConcentration}
Let $X$ be a random $n$-dimensional vector with independent entries
from $N(0,1)$. Then for $\epsilon >0$
\begin{align*}
\Pr\bigl(\norms{X} &\geq (1+\epsilon) n\bigr) \leq
\bigl((1+\epsilon)e^{-\epsilon}\bigr)^{n/2}
\end{align*}
and for $\epsilon \in (0,1)$
\begin{align*}
\Pr\bigl(\norms{X} &\leq (1-\epsilon) n\bigr) \leq
\bigl((1-\epsilon)e^{\epsilon}\bigr)^{n/2}.
\end{align*}
\end{lemma}
For a proof, see \cite[Lemma 1.3]{VemRP}.

\begin{lemma}\label{lem:randomBall}
Let $X$ be a uniform random vector in the $n$-dimensional ball of
radius $r$. Let $Y$ be an independent random $n$-dimensional unit
vector. Then,
\[
    \e(\norm{X}^{2}) = \frac{nr^2}{n+2}
\quad \mbox{ and }
\quad
    \E \bigl((\inner{X}{Y})^2\bigr) = \frac{r^2}{n+2}.
\]
\end{lemma}
\begin{proof}
For the first part, we have
\begin{align*}
    \e(\norm{X}^{2}) &= \frac{\int_0^r t^{n+1} dt}{\int_0^r t^{n-1}
    dt} = \frac{nr^2}{n+2}.
\end{align*}
For the second part, because of the independence and the symmetry we
can assume that $Y$ is any fixed vector, say $(1,0,\dotsc, 0)$. Then
$\E \bigl((\inner{X}{Y})^2\bigr) = \E (X_1^2)$. But
\[
\E (X_1^2) = \E (X_2^2) = \dotsb = \frac{1}{n} \sum_{i=1}^n
\e(X_i^2) = \frac{\E (\norm{X}^2)}{n} = \frac{r^2}{n+2}.
\]
\end{proof}

\begin{lemma}\label{lem:expectationVolume}
There exists a constant $c>0$ such that if $P \subseteq \RR^n$
compact and $X$ is a random point in $P$ then
\[
\E \norms{X} \geq c (\vol P)^{2/n} n
\]
\end{lemma}
%The previous lemma doesn't seem to improve if one also assumes that
%$P$ is a polytope with few facets: a cube  has $\E \norms{X} = O (
%(\vol P)^{2/n} n )$.
\begin{proof}
For a given value of $\vol P$, the value $\E \norms{X}$ is minimized
when $P$ is a ball centered at the origin. For some $c > 0$ we have
that the volume of the ball of radius $r$ is
\[
\frac{\pi^{n/2} r^n}{\Gamma(n/2+1)}
    = \frac{2\pi^{n/2} r^n}{n\Gamma(n/2)}
    \geq \frac{2 \pi^{n/2} r^n}{n (n/2)^{n/2}}
    \geq \frac{c^{n/2} r^n}{n^{n/2}}.
\]
This implies that, for a given value of $\vol P$, the radius $r$ of
the ball of that volume satisfies
\begin{equation}\label{equ:ballRadius}
\frac{c^{n/2} r^n}{n^{n/2}} \geq \vol P.
\end{equation}
On the other hand, Lemma \ref{lem:randomBall} claims that for $Y$ a
random point in the ball of radius $r$, we have
\begin{equation}\label{equ:ballExpectation}
\E \norms{Y}  = \frac{n r^2}{n+2}.
\end{equation}
Combining (\ref{equ:ballRadius}), (\ref{equ:ballExpectation}) and
the minimality of the ball, we get
\begin{align*}
\left(\frac{c \E \norms{X} (n+2) }{n^2}\right)^{n/2} \geq \vol P
\end{align*}
and this implies the desired inequality.
\end{proof}

We conclude this section with two elementary properties of variance.
\begin{lemma}\label{lem:productVariance} Let $X$, $Y$ be independent
real-valued random variables. Then
\[
\frac{\var (XY)}{(\E (XY))^2} = \left(1+ \frac{\var X}{(\e
X)^2}\right)\left(1+ \frac{\var Y}{(\E Y)^2}\right) - 1 \geq
\frac{\var X}{(\E X)^2} + \frac{\var Y}{(\E Y)^2}.
\]
\end{lemma}

\begin{lemma}\label{lem:SliceVar}
For real-valued random variables $X,Y$, $\var X = \e_Y \var(X
\giventhat Y) + \var_Y \e(X \giventhat Y)$.
\end{lemma}
%\begin{proof}
%\begin{align*}
%\var (XY) &= \E (X^2) \e(Y^2) - (\E X \E Y)^2 \\
%    &= [\var X + (\E X)^2] [\var Y + (\E Y)^2] - (\E X \E Y)^2 \\
%    &= (\E X)^2 (\E Y)^2 [1+ \frac{\var X}{(\E X)^2}][1+ \frac{\var Y}{(\E Y)^\
%2}] - (\E X \E Y)^2 \\
%    &= (\E (X Y))^2 [[1+ \frac{\var X}{(\E X)^2}][1+ \frac{\var Y}{(\E Y)^2}] \
%- 1]\\
%    &\geq (\E (X Y))^2 [\frac{\var X}{(\E X)^2} + \frac{\var Y}{(\E Y)^2}].
%\end{align*}
%\end{proof}

\section{Lower bound for length estimation}

In this section, we prove Theorem \ref{thm:lengthLowerBound}.
Let $a$ be uniform random vector from $[-1,1]^n$. By
Lemma \ref{lem:gaussianConcentration}, $\norm{a} \ge \sqrt{n}-4\sqrt{\log n}$ as required
by the theorem with probability
at least $1-1/n^2$. We will prove that
there exists a
constant $c>0$ such that any deterministic algorithm that outputs a
number $l$ such that
\[
\norm{a}-\frac{c}{\sqrt{\log n}} \leq l \leq
\norm{a}+\frac{c}{\sqrt{\log n}}
\]
with probability at least $1-O(1/n \log n)$ makes at least
$n-1$ halfspace queries. Along with Yao's lemma this proves the
theorem.

Our access to $a$ is via a membership oracle for the halfspace
$a\cdot x \le 1$.  Consider the decision tree of height $h$ for some
deterministic algorithm. This will be a binary tree. The
distribution at a leaf $l$ is uniform over the intersection of
$[-1,1]^n$ with the halfspaces given by the path (queries,
responses) to the leaf $l$ from the root $r$, i.e., uniform over a
polytope $P_l$ with at most $2n+h$ facets.

The volume of the initial set is $2^n$. The volume of leaves with
$\vol(P_l) < 1$ is less than $\card{L} = 2^h$ and so the total
volume of leaves with $\vol(P_l) \ge 1$ is at least $2^n-2^h$.
Setting $h=n-1$, this is $2^{n-1}$ and so with probability at least
$1/2$, $\vol(P_l) \ge 1$. For a random point $X$ from any such
$P_l$, Theorem \ref{VarPoly} implies that $\var \norms{X} \ge
cn/\log n$ for some absolute constant $c > 0$. Now by Lemma
\ref{lem:varImpliesDisp}(a), and the fact that the support of
$\norms{X}$ is an interval of length $n$, we get that for any
$b$,
\[
\Pr\left(\bigl\lvert\norms{X}-b\bigr\rvert \ge
\frac{1}{2}\sqrt{\frac{cn}{\log n}}\right) \ge \frac{3c}{4n\log n}.
\]
It follows that $\norm{X}$ is dispersed after $n-1$ queries. We note
that the lower bound can be extended to any algorithm that succeeds
with probability $1-1/n^{\eps}$ by a standard trick to boost the
success probability: we repeat the algorithm $O(1/\eps)$ times and
use the median of the results.

\section{Complexity of randomized volume algorithms}

We will use the distribution $D$ on parallelopipeds (or matrices,
equivalently). Recall that a random $n \times n$ matrix $R$ is
generated by choosing its rows $R_1, \dotsc, R_n$ uniformly and
independently from the ball of radius $\sqrt{n}$. The convex body
corresponding to $R$ is a parallelopiped having the rows of $R$ as
facets' normals:
\[
\{x \in \RR^n \suchthat (\forall i) \abs{\inner{R_i}{x}} \leq 1\}
\]
Its volume is $V: \RR^{n \times n} \to \RR$ given (a.s.) by $V(R) =
2^n \abs{\det R}^{-1}$.
%The multiplicative dispersions of
%$\abs{\det R}^{-1}$ and $\abs{\det R}$ are the same.

At a very high level, the main idea of the lower bound is the
following: after an algorithm makes all its queries, the set of
inputs consistent with those queries is a product set along rows (in
the oracle model $Q'$), while the level sets of the function that
the algorithm is trying to approximate, $\abs{\det(\cdot)}$, are far
from being product sets. In the partition of the set of inputs
induced by any decision tree of height $O(n^2/\log n)$, all parts
are product sets along rows and most parts have large volume, and
therefore $V$ is dispersed in most of them. To make this idea more
precise, we first examine the structure of a product set along rows
all with {\em exactly} the same determinant. This abstract
``hyperbola'' has a rather sparse structure.
% (proved in the appendix).

\begin{theorem}\label{thm:exact-hyperbola}
Let $R \subseteq \RR^{n \times n}$ be such that $R = \prod_{i=1}^n
R_i$, $R_i \subseteq \RR^n$ convex and there exists $c>0$ such that
$\abs{\det M } = c$ for all $M \in R$. Then, for some ordering of
the $R_i$'s, $R_i \subseteq S_i$, with $S_i$ an $(i-1)$-dimensional
affine subspace, $0 \notin S_i$ and satisfying: $S_i$ is a
translation of the linear hull of $S_{i-1}$.
\end{theorem}

\begin{proof}
By induction on $n$. It is clearly true for $n=1$. For arbitrary
$n$, consider the dimension of the affine hull of each $R_i$, and
let $R_1$ have minimum dimension. Let $a \in R_1$. There will be
two cases:

If $R_1=\{a\}$, then let $A$ be the hyperplane orthogonal to $a$. If
we denote $T_i$ the projection of $R_i$ onto $A$, then we have that
$T = \prod_{i=1}^{n-1} T_i$ satisfies the hypotheses in $A \cong
\RR^{n-1}$ with constant $c/\norm{a}$ and the inductive hypothesis
implies that, for some ordering, the $T_2, \dotsc, T_n$ are
contained in affine subspaces not containing 0 of dimensions $0,
\dotsc, n-2$ in $A$, that is, $R_2, \dotsc, R_n$ are contained in
affine subspaces not containing 0 of dimensions $1, \dotsc, n-1$.

If there are $a,b \in R_1$, $b \neq a$, then there is no
zero-dimensional $R_i$. Also, because of the condition on the
determinant, $b$ is not parallel to $a$. Let $x_\lambda = \lambda a
+ (1-\lambda) b$ and consider the argument of the previous paragraph
applied to $x_\lambda$ and its orthogonal hyperplane. That is, for
every $\lambda$ there is some region $T_i$ in $A$ that is
zero-dimensional. In other words, the corresponding $R_i$ is
contained in a line. Because there are only $n-1$ possible values of
$i$ but an infinite number of values of $\lambda$, we have that
there exists one region $R_i$ that is picked as the zero-dimensional
for at least two different values of $\lambda$. That is, $R_i$ is
contained in the intersection of two non-parallel lines, and it must
be zero-dimensional, which is a contradiction.
\end{proof}

Now we need to extend this to an approximate hyperbola, i.e., a
product set along rows with the property that for most of the
matrices in the set, the determinant is restricted in a given
interval. This extension is the heart of the proof and is captured
in Lemma \ref{thm:determinantDispersion}. We will need a bit of
preparation for its proof.

We define two properties of a matrix $R \in
\RR^{n \times n}$:
\begin{itemize}
\item Property $P_1(R, t)$: \( \prod_{i=1}^n
\norm{\proj_{R_{-i}^\perp}(R_i)} \leq t\) (``short 1-D
projections'').

%\item Property $P_1(R, \alpha)$: \( (\forall i \neq j)
%\norm{\proj_{R_{-ij}^\perp}(R_i)} \leq \alpha\) (``not too far in
%2-D projections'').
%
\item Property $P_2(R, t)$: \( \abs{\det \hat R } \geq t
\) (``angles not too small'').
\end{itemize}

\begin{lemma}\label{lem:goodProperties}
Let $R$ be drawn from distribution $D$. Then for any $\alpha > 1$,
\begin{itemize}

\item[a.] $\Pr\bigl(P_1(R, \alpha^n)\bigr) \geq 1-\frac{1}{\alpha^2},$

\item[b.] there exists $\beta>1$ (that depends on $\alpha$) such
    that $\Pr\bigl( P_2(R,1/\beta^n) \bigr) \geq 1 -
    \frac{1}{n^{\alpha}}$.

\end{itemize}
\end{lemma}

\begin{proof}
%Proof idea: for 1, we want to use Markov's inequality, but the
%expectation is not so easy to compute due to lack of independence.
%Thus, we turn the product into a sum by means of the AM-GM
%inequality.
%
For part (a), by the AM-GM inequality and Lemma \ref{lem:randomBall}
we have
\begin{align*}
\E \left(\Bigl(\prod_i
\norms{\proj_{R_{-i}^\perp}(R_i)}\Bigr)^{1/n}\right) &\leq
\frac{1}{n} \sum_i \E \norms{\proj_{R_{-i}^\perp}(R_i)} =
\frac{n}{n+2}.
\end{align*}
Thus, by Markov's inequality,
\begin{align*}
\Pr\left(\prod_i \norm{\proj_{R_{-i}^\perp}(R_i)} \geq c^n\right)
    &= \Pr\left(\Bigl(\prod_i \norms{\proj_{R_{-i}^\perp}(R_i)} \Bigr)^{1/n} \geq
  c^2\right) \leq \frac{1}{c^2}.
\end{align*}

%For 1, fix $i \neq j$. Then $\proj_{R_{-ij}^\perp}(R_i)$ is the
%projection of a random vector onto an independent random plane $P =
%R_{-ij}^\perp$. Without loss of generality we can assume that the
%plane is fixed. Let $u$, $v$ be an arbitrary basis of $P$. It is
%enough to bound the length of the projection of $R_i$ onto $u$ and
%$v$, as $\norm{\proj_P(R_i)} \leq \norm{\proj_u(R_i)} +
%\norm{\proj_v(R_i)}$. Now (Theorem ???, book3???),
%\begin{align*}
%\Prob(\norm{\proj_u(R_i)} > O(\sqrt{\log n})) \leq
%\frac{1-\alpha}{2n^2}.
%\end{align*}
%Thus, $\Prob(\norm{\proj_P(R_i)} > O(\sqrt{\log n})) \leq (1-p)/n^2$
%and $\Pr(\exists i,j, i\neq j) \norm{\proj_{R_{-ij}^\perp}(R_i)}
%> O(\sqrt{\log n})) \leq 1-\alpha$.

For part (b), we can equivalently pick each entry of $R$
independently as $N(0,1)$. In any case,
\[
\abs{\det \hat R} = \frac{\abs{\det R}}{\prod_i \norm{R_i}} = \frac{\prod_i
\norm{\tilde R_i}}{\prod_i \norm{R_i}}.
\]
We will find an upper
bound for the denominator and a lower bound for the numerator.

For the denominator, Markov's inequality and the fact that $\e \prod
\norms{R_i} = n^n$ give
\begin{equation}\label{equ:denominator}
\Pr\left(\prod_{i=1}^n \norms{R_i} \geq t n^n\right) \leq 1/t.
\end{equation}

For the numerator, let $\mu_i = \E \norms{\tilde R_i} = n-i+1$, let
$\mu = \E \prod_{i=1}^n \norms{\tilde R_i} = \prod_{i=1}^n \mu_i =
n!$.
%Then, for $\alpha_i >0$ that will be fixed later,
%\begin{align*}
%\Pr(\prod_{i=1}^n \norms{\tilde R_i} \geq \mu /\prod_{i=1}^n
%\alpha_i) &\geq \Pr((\forall
%i) \norms{\tilde R_i} \geq \mu_i/\alpha_i) \\
%    &= \prod_{i=1}^n \Pr( \norms{\tilde R_i} \geq \mu_i/\alpha_i ).
%\end{align*}
%Let $p(n) = 2/(1-\alpha)$.

Now, concentration of a Gaussian vector (Lemma
\ref{lem:gaussianConcentration}) gives
%\[
%\Pr( \norms{\tilde R_i} \geq t\mu_i ) \geq 1-
%e^{-(t-1)^2(n-i+1)/8}
%\]
\begin{equation}\label{equ:ineq1}
\Pr\bigl( \norms{\tilde R_i} \geq \mu_i/2 \bigr) \geq 1-
2^{-(n-i+1)/8}
\end{equation}
Alternatively, for $t \in (0,1)$ the fact that the density of
$N(0,1)$ is less 1 gives
\begin{equation}\label{equ:ineq2}
\Pr\bigl( \norms{\tilde R_i} \geq t\mu_i \bigr) \geq 1-
\sqrt{t}(n-i+1).
\end{equation}

%%%
%this
%implies that there exists a constant $c>0$ such that for $i \leq n -
%c\log n$ we have $\Pr( \norms{\tilde R_i} < \mu_i/2 ) \leq
%\frac{1}{n p(n)}$. For $i \leq n - c \log n$ we set $\alpha_i = 2$.
%On the other hand, for $i > n - c\log n$, we observe that for $X$ a
%$N(0,1)$ random variable, we have that $\Pr(\abs{X} \leq t) \leq
%t$.
%Using this, we set $\alpha_i = \mu_i (p(n) c n \log n)^2$ to get
%\begin{align*}
%\Pr(\norms{\tilde R_i} < \mu_i/\alpha_i) &\leq (n-i+1) \Pr(X^2 <
%\frac{\mu_i}{\alpha_i (n-i+1)}) \\
%&\leq (n-i+1) \Pr(X^2 < \frac{\mu_i}{\alpha_i}) \\
%&\leq c \log n\sqrt{\mu_i/\alpha_i} \\
%&= \frac{1}{n p(n)}.
%\end{align*}
Let $c>0$ be such that $2^{-(n-i+1)/8} \leq 1/(2n^{\alpha+1})$ for
$i \le n - c\log n$. Using inequality (\ref{equ:ineq1}) for $i \le n
- c\log n$ and (\ref{equ:ineq2}) for the
%rest with $t=t_2=1/n^{c+1}16\log n$,
rest with
\[
 t=\frac{1}{2n^{2\alpha} (c\log n)^{5/2}}
\]
we get
%\begin{align*}
%\Pr(\prod_{i=1}^n \norms{\tilde R_i} \geq \geq
%\frac{1}{\beta^{n}\mu}) &=
%\Pr(\prod_{i=1}^n \norms{\tilde R_i} \geq \geq t_2^{16c\log n} t_1^{n-16\log n}\mu) \\
%&= \prod_{i=1}^{n-16\log n} \Pr(\norms{\tilde{R_i}}  \geq \frac{\mu_i}{t_1})\prod_{i=n-16\log n}^{n} \Pr(\norms{\tilde{R_i}} \ge t_2\mu_i)\\
%&\geq 1- \frac{1}{n^{c+1}}.
%\end{align*}
\begin{equation}\label{equ:numerator}
\begin{aligned}
\Pr\biggl(\prod_{i=1}^n &\norms{\tilde R_i} \geq \frac{\mu}{2^{n -
c\log n} t^{c \log n}}\biggr) \\
&\geq \prod_{i=1}^{n-c\log n} \Pr\Bigl(\norms{\tilde{R_i}}  \geq \frac{\mu_i}{2}\Bigr) \prod_{i=n-c\log n}^{n} \Pr\bigl(\norms{\tilde{R_i}} \ge t\mu_i\bigr)\\
&\geq 1- \frac{1}{n^{\alpha}}
\end{aligned}
\end{equation}
where, for some $\gamma > 1$  we have $2^{n - c\log n} t^{c \log n}
\leq \gamma^n $. The result follows from equations
(\ref{equ:numerator}) and (\ref{equ:denominator}).
%%%% to be completed.
\killtext{ Also,
\begin{align*}
\prod_{i=1}^n \alpha_i &\leq 2^n \prod_{n-c \log n < i \leq n} \mu_i
(p(n)
cn \log n)^2 \\
&\leq 2^n (c \log n)^{3(c \log n)} (n p(n))^{2(c \log n)} \\
&= 2^{O(n)}
\end{align*}
and this implies
\begin{equation}\label{equ:numerator}
\begin{aligned}
\Pr(\prod_i \norms{\tilde R_i} \geq \mu /2^{O(n)}) &\geq
\Pr(\prod_i \norms{\tilde R_i} \geq \mu /\prod_i \alpha_i) \\
    &\geq 1 -1/p(n)
\end{aligned}
\end{equation}
}
%\begin{align*}
%\Pr(\det \hat R^2 \geq \frac{1}{2^{O(n)}}) &\geq \Pr(\det \hat
%R^2 \geq \frac{\mu}{4n^n 2^{O(n)}}) \\
%&\geq \Pr(\prod_i \norms{\tilde R_i} \geq \mu /2^{O(n)} \text{ and } \prod_i\
% \norms{R_i} < 4n^n) \\
%&\geq \alpha.
%\end{align*}
\end{proof}

\begin{proof}[Proof (of Lemma \ref{thm:determinantDispersion})]
The idea of the proof is the following: If we assume that
$\abs{\det(\cdot)}$ of most matrices in a part fits in an interval
$[u,u(1+\epsilon)]$, then for most choices $R_{-n}$ of the first
$n-1$ rows in that part we have that most choices $Y$ of the last
row in that part have $\abs{\det(R_{-n},Y)}$ in that interval. Thus,
in view of the formula\footnote{Recall that $\tilde R_i$ is the
projection of $R_i$ to the subspace orthogonal to $R_1, \dotsc,
R_{i-1}$.} $\abs{\det(R_{-n},Y)} = \norm{\tilde Y} \prod_{i=1}^{n-1}
\norm{\tilde R_i}$ we have that, for most values of $Y$,
\[
\norm{\tilde Y} \in \bigl[u,u(1 + \epsilon)\bigr] \prod_{i=1}^{n-1}
\norm{\tilde R_i}^{-1}
\]
where $\tilde Y$ is the projection of $Y$ to the line orthogonal to
$R_1, \dotsc, R_{n-1}$. In other words, most choices of the last row
are forced to be contained in a set of the form $\{ x \suchthat b
\leq \abs{\inner{a}{x}} \leq c \}$, that we call a double band, and
the same argument works for the other rows. In a similar way, we get
a pair of double bands of ``complementary'' widths for every pair of
rows. These constraints on the part imply that it has small volume,
giving a contradiction. This argument only works for parts
containing mostly ``matrices that are not too singular'' ---matrices
that satisfy $P_1$ and $P_2$---, and we choose the parameters of
these properties so that at least half of $(\sqrt{n}B_n)^n$
satisfies them.

We will firstly choose $N'$ as the family of large parts that
satisfy properties $P_1$ and $P_2$ for suitable parameters so that
(a) is satisfied. We will say ``probability of a subset of
$(\sqrt{n}B_n)^n$'' to mean its probability with respect to the
uniform probability measure on $(\sqrt{n}B_n)^n$. The total
probability of the parts having probability at most $\alpha$ is at
most $\alpha \card{N}$. Thus, setting $\alpha = 1/(4\card{N})$, the
parts having probability at least $1/4\card{N} \geq 1/2^{n^2}$ have
total probability at least $3/4$. Since $\vol \cup_{j \in N}
\mathcal{A}^j \geq 2^{n^2}$, each of those parts has volume at least
$1$. Let these parts be indexed by $N'' \subseteq N$. We choose
parameters in Lemma \ref{lem:goodProperties} (say, $\alpha = 4$ for
part (a), $\alpha = 2$ for part (b), giving the existence of some
$\beta$) so that at least 7/8 of $(\sqrt{n}B_n)^n$ satisfy
$P_1(\cdot, 4^n)$ and $P_2(\cdot, 1/\beta^n)$, and then at least
$3/4$ of the parts in probability satisfy $P_1(\cdot, 4^n)$ and
$P_2(\cdot, 1/\beta^n)$ for at least half of the part in
probability. Let $N''' \subseteq N$ be the set of indices of these
parts. Let $N' = N'' \cap N'''$. We have that $\cup_{j \in N'}
\mathcal{A}^j$ has probability at least $1/2$.
%Thus, at least half of the parts in volume satisfy the hypotheses of
%Proposition \ref{prop:volumeDispersion}. That is, there exists a
%constant $c>0$ such that with probability at least $1/(2^{17} n^4)$
%and for any $a
%> 0$ we have that $V(R)$ is outside of
%\[
%[a, (1+c)a].
%\]

We will now prove (b). Let $A=\prod_{i=1}^n A_i$ be one of the parts
indexed by $N'$. Let $X$ be random in $A$.
%Then there exists $c >
%0$ such that for any $u > 0$ and for $X \in A$ uniform
%\[
%\Pr( V(X) \notin [u,u(1+c)] ) \geq \frac{1}{2^{16} n^4}.
%\]
Let $\epsilon$ be a constant and $p_1(n)$ be a function of $n$ both to be
fixed later.
Assume for a contradiction
that there exists $u$ such that
\begin{equation}\label{equ:intervalContradiction}
\Pr\bigl( \abs{\det{X}} \notin [u,u(1+\epsilon)] \bigr) < p_1(n).
\end{equation}
Let $G \subseteq A$ be the set of $M \in A$ such that $\abs{\det{M}}
\in [u,u(1+\epsilon)]$. Let $p_2(n)$, $ p_3(n)$ be functions of $n$
to be chosen later. Consider the subset of points $R \in G$
satisfying:
\begin{itemize}
\item[I.] $P_1(R, 4^{n})$ and $P_2(R, 1/\beta^n)$,

\item[II.] for any $i \in \{1, \dotsc, n\}$, for at most a $p_2(n)$
fraction of $Y \in A_i$ we have $(Y, R_{-i}) \notin G$, and

\item[III.] for any $i,j \in \{1, \dotsc, n\}$, $i \neq j$, for at most a $p_3(n)$ fraction of $(Y, Z)
\in A_i \times A_j$ we have $(Y, Z, R_{-ij}) \notin G$.
\end{itemize}
Because of the constraints, such a subset is a
\begin{multline}\label{equ:choiceOfP1}
1 - \Pr(X \notin G) - \Pr(X \in G \text{ and not as I, II and III}) \geq \\
\geq 1 -
p_1(n) - \frac{1}{2} - n\frac{p_1(n)}{p_2(n)} - n^2 \frac{
p_1(n)}{p_3(n)}
\end{multline}
fraction of $A$. The function $p_1(n)$ will be chosen at the end so
that the right hand side is positive. Fix a matrix $R = (R_1,
\dotsc, R_n)$ in that subset.

%The idea of the rest of the proof is based on the following
%observation: if we assume that $\abs{\det{(\cdot)}}$ of most points
%in $A$ fits in a small interval (Equation
%(\ref{equ:intervalContradiction})), then each $A_i$ is restricted to
%certain ``bands'', and their intersection gives an upper bound for
%the volume of $A$: First, in view of the formula\footnote{Recall
%that $\tilde R_i$ is the projection of $R_i$ to the subspace
%orthogonal to $R_1, \dotsc, R_{i-1}$.} $\abs{\det R} = \prod_{i=1}^n
%\norm{\tilde R_i}$ and property (b) of $R$, we have that for $Y \in
%A_n$ except for at most a $p_2(n)$ fraction:
%\[
%\norm{\tilde Y} \in \bigl[u,u(1 + \epsilon)\bigr] \prod_{i=1}^{n-1}
%\norm{\tilde R_i}^{-1}
%\]
%where $\tilde Y$ is the projection of $Y$ to the line orthogonal to
%$R_1, \dotsc, R_{n-1}$. In other words, most of $A_n$ is forced to
%be contained in a set of the form $\{ x \suchthat b \leq
%\abs{\inner{a}{x}} \leq c \}$, that we call a ``band'', and the same
%argument works for the other $A_i$'s. In a similar way, property (c)
%gives a pair of bands of ``complementary'' widths for every pair
%$i,j$. These constraints on $A$ imply that it has small volume,
%giving a contradiction.
%%convex case
%The parallelopiped for $A_i$ is the intersection of all the
%corresponding bands.

The constraints described in the first paragraph of the proof are
formalized in Lemma \ref{lem:2Dlemma}, which, for all $i$, $j$,
gives sets $B_{ij}$ (double bands, of the form $\{ x \suchthat b
\leq \abs{\inner{a}{x}} \leq c \}$), such that most of $A_i$ is
contained in $ \cap_{j=1}^n B_{ij} $. Lemma \ref{lem:2Dlemma} is
invoked in the following way: For each pair $i,j$ with $i < j$, let
$E$ be the two-dimensional subspace orthogonal to all the rows of
$R$ except $i, j$. We set $X_1$ (respectively $X_2$)
%as the marginal onto $R_{-ij}^\perp$ of the uniform probability
%measure on $A_i$ (respectively $A_j$)\footnote{That is,
%$$\mu_1(\mathcal{A}) = V(A_i \cap
%\proj^{-1}_{R_{-ij}^\perp}(\mathcal{A}) )/V(A_i).$$},
distributed as the marginal in $E$ of the uniform probability
measure on $A_i$ (respectively $A_j$).
%% convex case
%These marginals are logconcave and have bounded support.
We also set $a_1=\proj_{E}(R_i)$, $a_2=\proj_{E}(R_j)$, $p =
p_3(n)$, $q = p_2(n)$ and $u$ and $\epsilon$ as here, while $\gamma$
will be chosen later.

Let $l_{ij}$ be the width of (each component of) the double band
$B_{ij}$. Then, according to Lemma \ref{lem:2Dlemma}, the following
relations hold:
\begin{align*}
l_{ii} &\leq \epsilon \norm{\proj_{R_{-i}^\perp}(R_i)} & \text{for any $i$,} \\
l_{ij} &\leq 4 \epsilon
\norm{\proj_{R_{-i}^\perp}(R_i)}\norm{\proj_{R_{-j}^\perp}(R_j)}/
l_{ji} & \text{for $i > j$.}
\end{align*}

%To see the bound on the volume of $A$ it is more convenient to
%arrange these values in a matrix having the width of $B_{ij}$ as the
%entry $i,j$:
%\[
%\begin{pmatrix}
%\epsilon \norm{\proj_{R_{-1}^\perp}(R_1)} & l_{12} & \cdots & l_{1n}
%\\
%4 \epsilon
%\norm{\proj_{R_{-1}^\perp}(R_1)}\norm{\proj_{R_{-2}^\perp}(R_2)}/
%l_{12} & \epsilon \norm{\proj_{R_{-2}^\perp}(R_2)} \\
%\vdots & & \ddots &\\
%4 \epsilon
%\norm{\proj_{R_{-1}^\perp}(R_1)}\norm{\proj_{R_{-n}^\perp}(R_n)}/
%l_{1n} & & &\epsilon \norm{\proj_{R_{-n}^\perp}(R_n)}
%\end{pmatrix}
%\]
%% non-convex case
Since each double band has two components, the intersection of all
the $n$ bands associated to a particular region $A_i$, namely
$\cap_{j=1}^n B_{ij}$, is the union of $2^n$ congruent
parallelopipeds. Thus, using properties $P_1$ and $P_2$ of $R$ and
fixing $\epsilon$ as a sufficiently small constant, the ``feasible
region'' defined by the double bands, $B = \prod_{i=1}^n
\cap_{j=1}^n B_{ij}$, satisfies:
%% non-convex case
\begin{align*}
\vol B &\leq 2^{n^2} \frac{\prod_{i,j=1}^n l_{ij}}{\abs{\det{\hat
R}}^n} \\
&\leq 2^{n^2} \frac{\prod_{i=1}^n \left(\epsilon
\norm{\proj_{R_{-i}^\perp}(R_i)} \prod_{j=2}^i 4 \epsilon
\norm{\proj_{R_{-i}^\perp}(R_i)}\norm{\proj_{R_{-j}^\perp}(R_j)}\right)}{\abs{\det{\hat
R}}^n} \\
&= 2^{n^2} \frac{\epsilon^{\binom{n}{2}} 4^{\binom{n-1}{2}} \prod_i
\norm{\proj_{R_{-i}^\perp}(R_i)}^n }{\abs{\det{\hat R}}^n}  \\
%&\leq 2^{O(n^2)} \epsilon^{\binom{n}{2}}
&\leq 1/4^n.
\end{align*}
%% convex case
%\begin{align*}
%\vol B &\leq \frac{\prod_{i=1}^n (\epsilon
%\norm{\proj_{R_{-i}^\perp}(R_i)} \prod_{j=2}^i 8 \epsilon
%\norm{\proj_{R_{-i}^\perp}(R_i)}\norm{\proj_{R_{-j}^\perp}(R_j)})}{\abs{\det{\hat
%R}}^n} \\
%&= \frac{\epsilon^{\binom{n}{2}} 8^{\binom{n-1}{2}} \prod_{i=1}^n
%\norm{\proj_{R_{-i}^\perp}(R_i)}^n }{\abs{\det{\hat R}}^n}  \\
%%&\leq 2^{O(n^2)} \epsilon^{\binom{n}{2}} \\
%&\leq 1/4^n.
%\end{align*}
Each region $A_i$ is not much bigger than the intersection of the
corresponding double bands $B_i = \cap_{j=1}^n B_{ij}$ as follows:
restricting to the double band $B_{ii}$ removes at most a $p_2(n)$
fraction of $A_i$, each double band $B_{ij}$ for $j < i$ removes at
most a $\gamma$ fraction of $A_i$, and each double band $B_{ij}$ for
$j > i$ removes a $p_2(n)+ (p_3(n)/\gamma)$ fraction of $A_i$. We
set $\gamma = 1/4 n^2 $, $p_2(n) = 1/(4 n^2)$ and $p_3(n) = 1/(16
n^4)$ so that, as a fraction of $\vol A_i$, $\vol B_i$ is no less
than
\begin{align*}
1 - n p_2(n) - \binom{n}{2} \gamma - \binom{n}{2}\left(p_2(n)+
\frac{p_3(n)}{\gamma}\right)
    &\geq 1/2.
\end{align*}
Thus, $\vol A \leq 2^n \vol B \leq 1/2^n$, which is a contradiction.
The condition on $p_1(n)$ given by Equation (\ref{equ:choiceOfP1})
is satisfied for $p_1(n) = 1/(2^7 n^6)$.
\end{proof}

%For the next lemma we need the following notation: for $i \in \{1,
%2\}$, let $\bar i = 1$ if $i=2$ and $\bar i= 2$ if $i=1$.
\begin{lemma}[2-D lemma]\label{lem:2Dlemma}
Let $X_1, X_2$ be two independent random vectors in $\RR^2$ with
bounded support (not necessarily with the same distribution). Let
$X$ be a random matrix with rows $X_1, X_2$. Assume that there exist
$u>0$, $0 <\epsilon \leq 1$ such that
\[
\Pr\bigl(\abs{\det{X}} \notin [u,u(1+\epsilon)]\bigr) < p. %(p_3)
\]
Let $G = \{M  \in \RR^{2 \times 2} \suchthat \abs{\det{M}} \in
[u,u(1+\epsilon)]\}$. Let $a_1, a_2 \in \RR^2$ be such that $(a_1,
a_2) \in G$ and
\[
\Pr(X_1 \suchthat (X_1, a_2) \notin G) \leq q, %(p_2)
\qquad
\Pr(X_2 \suchthat (X_2, a_1) \notin G) \leq q. %(p_2)
\]
Let $\gamma > p/(1-q)$. Then there exist double bands $B_{ij}
\subseteq \RR^2$, $b_{ij} \geq 0$, $i,j \in \{1,2\}$, $l \geq 0$,
\begin{align*}
B_{11} &= \Bigl\{ x \suchthat \abs{\inner{a_2^\perp}{x}} \in
\bigl[b_{11}, b_{11}+
\epsilon\norm{\pi_{a_2^\perp}(a_1)}\bigr]\Bigr\} \\
B_{22} &= \Bigl\{ x \suchthat \abs{\inner{a_1^\perp}{x}} \in
\bigl[b_{22}, b_{22}+
\epsilon\norm{\pi_{a_1^\perp}(a_2)}\bigr]\Bigr\} \\
B_{12} &= \Bigl\{ x \suchthat \abs{\inner{a_1^\perp}{x}} \in
\bigl[b_{12},
b_{12}+ l\bigr]\Bigr\} \\
B_{21} &= \Bigl\{ x \suchthat \abs{\inner{a_2^\perp}{x}} \in
\bigl[b_{21}, b_{21}+ 4 \epsilon
\norm{\proj_{a_2^\perp}(a_1)}\norm{\proj_{a_1^\perp}(a_2)} /
l\bigr]\Bigr\}
\end{align*}
such that
\begin{align*}
\Pr(X_1 \notin B_{11}) &\leq q &
\Pr(X_1 \notin B_{12}) &\leq q + (p/\gamma) \\
\Pr(X_2 \notin B_{21}) &\leq \gamma & \Pr(X_2 \notin B_{22}) &\leq
q.
\end{align*}
%% logconcave case
%If in addition $\mu_1, \mu_2$ have logconcave densities, then one
%can remove the absolute values in the definition of the bands while
%making the bounds on the probabilities $3/2$ times larger (i.e.
%$B_{11} = \{ x \suchthat \inner{a_2^\perp}{x} \in [b_{11}, b_{11}+
%\epsilon\norm{\pi_{a_2^\perp}(a_1)}]\}$, $\Pr(a_1' \notin B_{11})
%\leq 3\beta/2$ and similarly for the others).
\end{lemma}

\begin{proof}
The proof refers to Figure \ref{fig:twodim} which depicts the bands
under consideration.

\begin{figure}%[htbp]
\begin{center}
%\resizebox{.7\columnwidth}{!}{\input{2Dlemma.pstex_t}}
%\includegraphics[width=.7\columnwidth]{2dlemma.eps}
\includegraphics[width=.7\columnwidth]{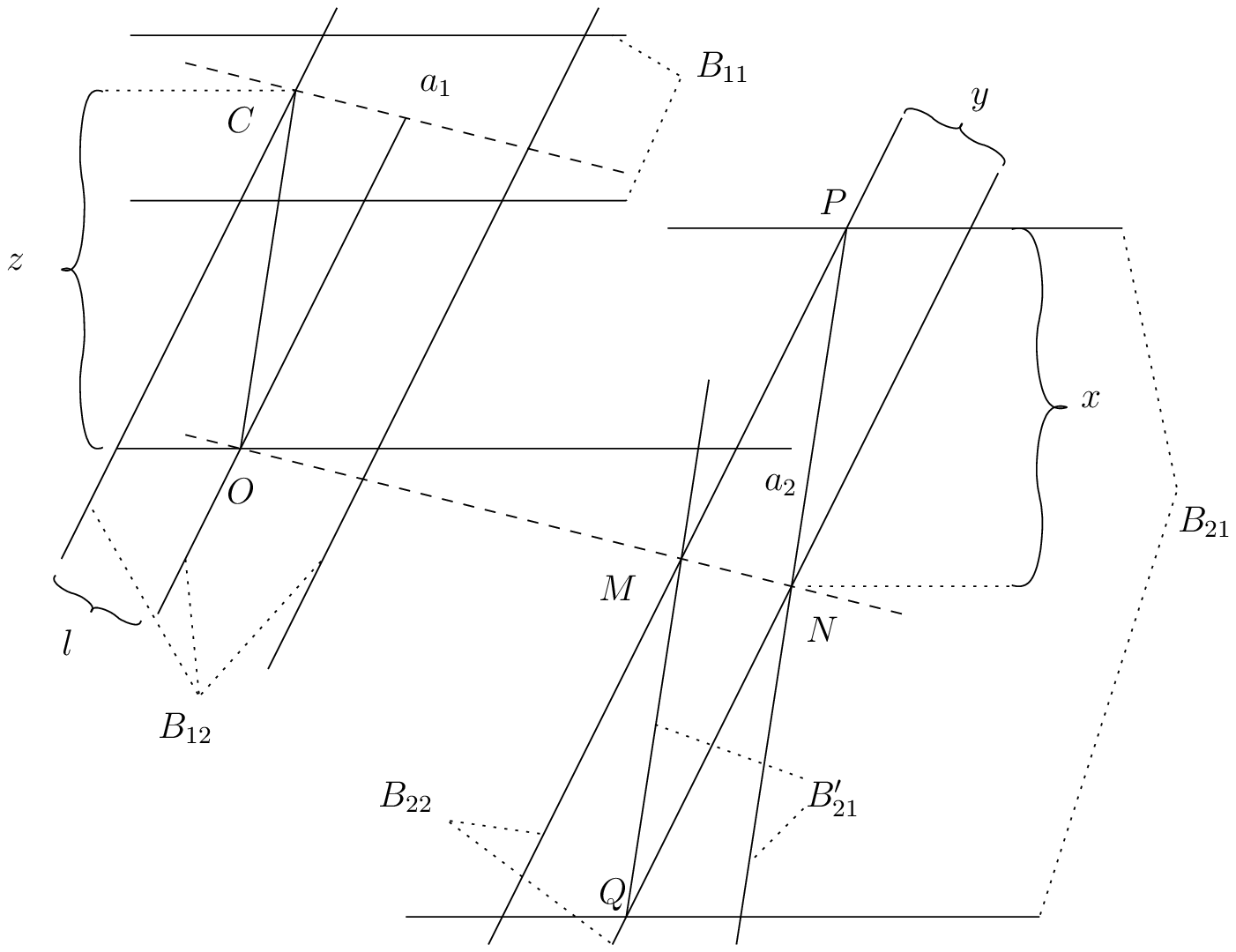}
    \caption{The 2-D argument.}\label{fig:twodim}
\end{center}
\end{figure}

A double band of the form $\{x \suchthat \abs{\inner{a}{x}} \in
[u,v]\}$ has (additive or absolute) width $v-u$ and relative (or
multiplicative) width $v/u$. Consider the expansion $\abs{\det X} =
\norm{X_2}\norm{\proj_{X_2^\perp}(X_1)}$ and the definition of $a_2$
to get
\[
\Pr\bigl(\norm{\proj_{a_2^\perp}(X_1)} \notin
\norm{a_2}^{-1}[u,u(1+\epsilon)]\bigr) \leq q.
\]
That is, with probability at most $q$ we have $X_1$ outside of a
double band of relative width $1+\epsilon$:
\[
B_{11} = \bigl\{ x \suchthat  \norm{\proj_{a_2^\perp}(x)} \in
\norm{a_2}^{-1}[u,u(1+\epsilon)]\bigr\}.
\]
Because $a_1 \in B_{11}$, the absolute width is at most $\epsilon
\norm{\proj_{a_2^\perp}(a_1)}$. If we exchange the roles of $a_1$
and $a_2$ in the previous argument, we get a double band $B_{22}$.

%For any $i$, let $A_i'$ be the set of points $R_i' \in A_i$
%satisfying: For any $j \neq i$ and for at most a $1/q_2(n)$ fraction
%of $R_{j} \in A_j$ we have $(R_j, R_i', R_{-ij}) \notin G$.
%
%Fix $i$, $j$ and focus on the projection of $A_i$ and $A_j$ to the
%plane orthogonal to $R_{-ij}$. Consider a point $D \in A_i'$ that
%maximizes the distance to $R_i$ in the direction orthogonal to
%$R_i$, that is, it maximizes $\abs{\proj_{R_{-j}^\perp}(\cdot)}$. By
%definition of $A_i'$, at most a $1/q_2(n)$ fraction of $

%Let $\gamma = \sqrt{p}$.
Let $\mathcal{A}$ be the set of $a \in \RR^2$ satisfying: $(a,a_2)
\in G$ and with probability at most $\gamma$ over $X_2$ we have $(a,
X_2) \notin G$. We have that $$\Pr(X_1 \in \mathcal{A}) \geq
1-q-\frac{p}{\gamma}.$$ Consider a point $C \in \mathcal{A}$ that
maximizes
%%\comment{$\mathcal{A}$ bounded, but not necessarily closed}
the distance to the span of $a_1$. Similarly to the construction of
$B_{11}$, by definition of $\mathcal{A}$ and with probability at
most $\gamma$ we have $X_2$ outside of a double band of relative
width $1+\epsilon$. We denote it $B_{21}'$. In order to have better
control of the angles between the bands, we want to consider a
bigger double band parallel to $B_{11}$, the minimum such a band
that contains the intersection of $B_{22}$ and $B_{21}'$. Call this
band $B_{21}$. Consider the line though the origin $O$ parallel to
$C-a_1$, and points $M$ and $N$ where the boundary of one component
of the double band $B_{22}$ intersects the line, $M$ is the point
closest to the origin, $N$, the farthest. The boundary of $B_{21}'$
intersects the boundary of $B_{11}$ precisely at $\pm M$ and $\pm
N$, because for any vector $v \in \RR^2$ parallel to $C-a_1$ we have
$\abs{\det(v,C)} = \abs{\det(v,a_1)}$. Consider the components of
$B_{21}'$ and $B_{22}$ containing $M$ and $N$ and let $P$ be any of
the other two points where the boundaries of those components meet.
This implies that triangles $Oa_1C$ and $PMN$ are similar. The width
of $B_{21}$ is at most $2x$, where $x = \max \{
\norm{\proj_{a_2^\perp}(P-M)}, \norm{\proj_{a_2^\perp}(P-N)}\}$.
Then,
\[
\frac{x}{z} = \frac{y}{l},
\]
where $l = \norm{\proj_{a_1^\perp}(C )}$ is the width of a band
imposed on $\mathcal{A}$ by definition of $C$, $y$ is the width of
$B_{22}$, $y \leq \epsilon \norm{\proj_{a_1^\perp}(a_2)}$, and $z$
is the distance between $C$ or $a_1$ and the span of $a_2$,
whichever is larger, that is,
\[z =
\max \{ \norm{\proj_{a_2^\perp}(C)}, \norm{\proj_{a_2^\perp}(a_1)}\} \leq (1 + \epsilon)
\norm{\proj_{a_2^\perp}(a_1)} \leq 2
\norm{\proj_{a_2^\perp}(a_1)}.
\]
Thus, $x \leq 2 \epsilon
\norm{\proj_{a_2^\perp}(a_1)}\norm{\proj_{a_1^\perp}(a_2)} / l$. Let
$B_{12}$ be the band imposed on $\mathcal{A}$ by definition of $C$.
%% logconcave case
%In the logconcave case, we analyze each band independently. Consider
%for example $B_{11}$. The marginal of $\mu_1$ onto the direction of
%the band, $a_i^\perp$, has a one-dimensional logconcave density $f$,
%in particular, $f$ is unimodal. Using this, we will prove that one
%of the components of $B_{11}$ has probability at most $q/2$, and
%we can just remove that component from the band. To see this, assume
%for a contradiction that both components have probability more than
%$q/2$. Without loss of generality $b_{11}=1$. Then, there exist
%$t_+ \geq 1$, $p_- \leq -1$ such that $f(t_+) >
%q/(2\epsilon)$, $f(t_-) > q/(2\epsilon)$. The unimodality of
%$f$ implies that $f(x) > q/(2\epsilon)$ for $x \in [-1,1]$ and
%$\nu([-1,1]) > q / \epsilon \geq q$, which is a
%contradiction.
\end{proof}

We are now ready to prove the complexity lower bounds.
\begin{proof}[Proof of Theorem \ref{thm:determinantLowerBound}.]
In view of Yao's lemma, it is enough to prove a lower bound on the
complexity of deterministic algorithms against a distribution and
then a lower bound on the minimum singular value of matrices
according to that distribution. The deterministic lower bound is a
consequence of the dispersion of the determinant proved in Lemma
\ref{thm:determinantDispersion}, the bound on the minimum singular
value is an easy adaptation of a bound on the minimum singular value
of a Gaussian matrix given by Lemma \ref{MINSINGULAR}. These two
claims are formalized below.

\emph{Claim 1:} Let $R$ be a random input according to distribution
$D$. Then there exists a constant $c>0$ such that any deterministic
algorithm that outputs a number $V$ such that
\[
(1-c)\abs{\det{R}} \leq V \leq (1+c)\abs{\det{R}}
\]
with probability at least $1-1/( 2^{8} n^6)$ makes more than
\[
\frac{n^2 - 2}{\log_2 (2n+1)}
\]
queries in the oracle model $Q'$.

\emph{Claim 2:} Let $A$ be an $n \times n$ random matrix from
distribution $D$. Let $\sigma$ be the minimum singular value of $A$.
Then for any $t \geq 0$
\[
\Pr(\sigma \sqrt{n} \leq t) \leq 4t + \frac{n}{2^{n-1}}
\]
(the choice of $t=1/(2^{12} n^6)$ proves Theorem
\ref{thm:determinantLowerBound}).
% additional factor of 1/2 to take
%into account lower order term in Lemma \ref{lem:minSingularBall}

\emph{Proof of Claim 1:} For a deterministic algorithm and a value
of $n$, consider the corresponding decision tree. Let
\[
    h \leq \frac{n^2 - 2}{\log_2 (2n+1)}
\]
be the height and $L$ be the set of
leaves of this tree. Let $(P_l)_{l \in L}$ be the partition on the
support of $D$ induced by the tree.

Every query has at most $2n + 1$ different answers, and every path
has height at most $h$. Thus,
\[
\card{L} \leq (2n+1)^h = 2^{n^2 - 2}.
\]
%The total probability of the leaves having probability at most
%$\alpha$ is at most $\alpha \card{L}$. Thus, setting $\alpha =
%1/(4\card{L})$, the leaves having probability at least
%\[
%\frac{1}{4\card{L}} \geq \frac{1}{2^{n^2}}
%\]
%have total probability at least $3/4$. Because $\vol \cup_{l \in L}
%P_l \geq 2^{n^2}$, we have that each of those leaves has volume at
%least $1$.
%
%Moreover, Lemma \ref{lem:goodProperties} (with $\alpha = 7/8$, $c =
%\sqrt{8}$) implies that at least $3/4$ of the leaves in volume
%satisfy $P_1(\cdot, 8^{n/2}))$ and $P_2(\cdot, 1/\beta^n)$ for at
%least half of the leaf in volume. Thus, at least half of the leaves
%in volume satisfy the hypotheses of Proposition
%\ref{prop:volumeDispersion}. That is, there exists a constant $c>0$
%such that with probability at least $1/(2^{17} n^4)$ and for any $a
%> 0$ we have that $V(R)$ is outside of
%\[
%[a, (1+c)a].
%\]
The sets $P_l$ are product sets along rows by Lemma \ref{LEM:TREE},
and hence by Lemma \ref{thm:determinantDispersion} we have that
there exists a constant $c>0$ such that with probability at least
$1/(2^{8} n^6)$ and for any $a > 0$ we have that $\abs{\det{R}}$ is
outside of $[a, (1+c)a]$. Claim 1 follows.

\emph{Proof of Claim 2:} We will bound
$\norm{A^{-1}}_2 = 1/\sigma$. To
achieve this, we will reduce the problem to the case where the
entries of the matrix are $N(0,1)$ and independent.
We write $A = G D E$, where $G$ has its entries
independently as $N(0,1)$, $D$ is the diagonal matrix that
normalizes the rows of $G$ and $E$ is another random diagonal matrix
independent of $ (G, D)$ that scales the rows of $G D$ to give them
the length distribution of a random vector in $\sqrt{n} B_n$. We
have
\begin{equation}\label{equ:2norm}
\norm{A^{-1}}_2 \leq \norm{D^{-1}}_2 \norm{E^{-1}}_2
\norm{G^{-1}}_2.
\end{equation}
Now, with probability at least $ 1 - n / 2^n$ the diagonal entries
of $E$ are at least $\sqrt{n}/2$. Thus, except for an event that
happens with probability $n /2^n$,
\begin{equation}\label{equ:event1}
\norm{E^{-1}}_2 \leq 2/\sqrt{n}.
\end{equation}
On the other hand, Lemma \ref{lem:gaussianConcentration} (with
$\epsilon=3$) implies that with probability at least $1- n/2^n$ the
diagonal entries of $D^{-1}$ are at most $2\sqrt{n}$. Thus, except
for an event that happens with probability $n/2^n$,
\begin{equation}\label{equ:event2}
\norm{D^{-1}}_2 \leq 2 \sqrt{n}.
\end{equation}

{}From (\ref{equ:2norm}), (\ref{equ:event1}) and (\ref{equ:event2}),
we get $\norm{A^{-1}}_2 \leq 4\norm{E^{-1}}$. Using Lemma
\ref{MINSINGULAR} which bounds the singular values for a Gaussian
matrix, Claim 2 follows.
\end{proof}

Finally, Theorem \ref{thm:volumeLowerBound} is a simple consequence.
\begin{proof}[Proof of Theorem \ref{thm:volumeLowerBound}.]
It remains to prove that a parallelopiped given by a matrix $A$ as
in Theorem \ref{thm:determinantLowerBound} contains $B_n/\sqrt{n}$
and is contained in $\frac{\sqrt{n}}{\sigma} B_n$ whenever $\sigma >
0$, where $\sigma$ is the minimum singular value of $A$. The first
inclusion is evident since the entries must be from $[-1,1]$. It is
sufficient to prove the second inclusion for the vertices of the
parallelopiped, i.e., solutions to $A x = b$ for any $b \in \{-1,
1\}^n$. That is, $x = A^{-1} b$ and therefore
\begin{equation*}
\norm{x} \leq \norm{A^{-1}}_2 \norm{b} \leq \sqrt{n}/ \sigma.
\end{equation*}
\end{proof}

\subsection{Nonadaptive volume algorithms}\label{nonadaptive}
An algorithm is \emph{nonadaptive} if its queries
are independent of the input.
\begin{theorem}[nonadaptive lower bound]\label{thm:nonadaptive}
Let $K$ be a convex body given by a membership oracle such that $B_n
\subseteq K \subseteq 2nB_n$. Then any nonadaptive randomized
algorithm that outputs a number $V$ such that $.9 \vol(K) \leq V
\leq 1.1 \vol(K)$ holds with probability at least $3/4$  has
complexity at least $\frac{1}{2 e (n+2)} n^{n/2}$.
\end{theorem}
\begin{proof}
Consider the distribution on parallelopipeds induced by the
following procedure: first, with equal probability choose one of the
following bodies:
\begin{itemize}
\item (``brick'') $\bigl\{ x \in \RR^n \suchthat (\forall i \in \{2, \dotsc, n \}) \
\abs{x_i} \leq 1\bigr\} \cap nB_n$
\item (``double brick'') $\bigl\{ x \in \RR^n \suchthat (\forall i \in \{2, \dotsc,\
 n \})\ \abs{x_i} \leq 1\bigr\} \cap 2nB_n$
\end{itemize}
and then, independently of the first choice, apply a random
rotation.

We will prove the following claim, from which the desired conclusion
can be obtained by means of Yao's lemma.

{\em Claim:} Let $K$ be a parallelopiped according to the previous
distribution. Then any nonadaptive deterministic algorithm that
outputs a number $V$ such that
\begin{equation}\label{equ:nonadaptive}
.9\vol(K) \leq V \leq 1.1\vol(K)
\end{equation}
holds with probability more than $\frac{1}{2} + \frac{Qn}{2}
(\frac{2}{n \pi})^{n/2}$ has complexity at least $Q$.

{\em Proof of Claim:} To satisfy Equation (\ref{equ:nonadaptive}),
the algorithm has to actually distinguish between the brick and the
double brick.
%Let
%the \emph{bad facets} be the two facets of the brick orthogonal to
%the first coordinate before the rotation.
Let the \emph{bad surface} be the intersection between the input and
the sphere of radius $n$. In order to distinguish between the two
bodies, the algorithm has to make at least one query whose ray hits
the bad surface. We will prove that the probability of this event is
no more than $ 2Q (2/e\pi n)^{n/2}$. To see this, observe that the
probability of a query hitting the bad surface is at most the volume
of the bad surface divided by the volume of the sphere of radius
$n$. The former can be bounded in the following way: Let $x = (x_2,
\dotsc, x_n)$ be the coordinates along the normals to the $n-1$
facets of the body. Parameterize one of the hemispheres determined
by the hyperplane containing those normals as $F(x_2, \dotsc, x_n) =
\sqrt{n^2 - x_2^2 - \dotsb - x_n^2}$.

We have that
\[
    \frac{d}{dx_i} F(x) = \frac{x_i}{F(x)}.
\]
In the domain of integration $[-1,1]^{n-1}$ we have $\norm{x}^2 \leq
n$ and this implies that in that domain
\[
\norm{\grad F(x)}^2 = \frac{\norm{x}^2}{n^2 - \norm{x}^2} \leq
\frac{1}{n-1}.
\]
The volume of the bad surface is given by
%\comment{check??}
\[
   2 \int_{[-1,1]^{n-1}} \sqrt{1+\norm{\grad F(x)}^2} \; dx
 \leq 2^{n} \sqrt{1 + \frac{1}{n-1}} \leq 2^{n+1}
\]
The volume of the sphere of radius $n$ is
\[
    \frac{2n^{n-1} \pi^{n/2}}{\Gamma(n/2)} \geq
    \frac{2n^{n-1} \pi^{n/2}}{(n/2)^{n/2}}
    = \frac{2}{n} (2 n \pi)^{n/2}.
\]
Thus, the probability that a particular query hits the bad surface
is at most
\[
n \left(\frac{2}{n \pi}\right)^{n/2}.
\]

Therefore the algorithm gives the wrong answer with probability at
least
\[
    \frac{1}{2} \left(1 - Q n \left(\frac{2}{n \pi}\right)^{n/2}\right).
\]
\end{proof}

\section{Lower bound for the product}\label{sec:product}

\begin{proof} (of Lemma \ref{UNIFORMPART}.)
Let the distribution function be
$F(t) = \Pr(X \le t) = e^{g(t)}$ for some concave
function $g$ and the density is $f(t) = g'(t)e^{g(t)}$ where $g'(t)$ is
nonincreasing.
First, we observe that logconcavity implies that
$F(\mu) \ge 1/4$. To see this, let $\mu-l$ be the point where $F(\mu-l) = F(\mu)/2$. Then, $F(\mu-il) \le F(\mu)/2^i$ and
\begin{align*}
\int_{0}^\mu (\mu-x) f(x) \, dx &\le \sum_{i \ge 1} \bigl(F(\mu-(i-1)l)-F(\mu-il)\bigr)(il)\\
&\le F(\mu)l + \sum_{i > 1}F(\mu-il)\bigl((i+1) - i\bigr)l\\
&\le F(\mu)l\sum_{i \ge 0} \frac{1}{2^i} = 2lF(\mu).
\end{align*}
On the other hand (assuming $F(\mu) \le 1/4$, otherwise, there is
nothing to prove),
\[
\int_\mu^{\infty} (x-\mu)f(x)\, dx \ge \sum_{i=
1}^{\lfloor\log(1/F(\mu))\rfloor} (2^i-2^{i-1})F(\mu)(i-1)l \ge
\frac{\log\bigl(1/F(\mu)\bigr)}{2} l.
\]
Therefore, we must have $2F(\mu) \ge \log(1/F(\mu))/2$ which implies
$F(\mu) \ge 1/4$.

%In the interval $I=[\mu-l,\mu]$, for any $t \in I$,
%\[
%\frac{f(t)}{f(\mu)} = \frac{g'(t)F(t)}{g'(\mu)F(\mu)} \ge \frac{1}{2}.
%\]
%If $l \ge \sigma/10$ and
%$f(\mu) \ge 1/10l$, then we
%set $[a,b]=I$. The density in this interval is at least $1/20l
%and we have $\alpha \ge 1/80$ proving the lemma.
Next,
\[
\int_0^\mu (\mu-x)f(x)\,dx \ge \int_0^{\mu-l} (\mu-x)f(x)\, dx
\ge F(\mu-l)l \ge \frac{l}{8}.
\]
Therefore, since $\mu$ is the mean,
\[
\int_\mu^{\infty}(x-\mu)f(x)\, dx \ge \frac{l}{8}.
\]
It follows that
\begin{equation}\label{right}
\int_\mu^{\infty} (x-\mu)^2 f(x)\, dx \ge \frac{l^2}{64}.
\end{equation}

Suppose $l < \sigma/4$.
Then,
\begin{align*}
\int_{0}^\mu (x-\mu)^2f(x)\,dx
&\le  \sum_{i \ge 1} \bigl(F(\mu-(i-1)l)-F(\mu-il)\bigr)(il)^2\\
&\le F(\mu)l^2 + \sum_{i > 1}F(\mu-il)\bigl((i+1)^2 - i^2\bigr)l^2\\
&\le F(\mu)l^2\sum_{i \ge 1} \frac{2i+1}{2^i} = 5l^2F(\mu) \le \sigma^2/2.
\end{align*}
Since
\[
\sigma^2 = \int_{0}^\infty (x-\mu)^2 f(x) \, dx = \int_0^\mu (x-\mu)^2 f(x)\, dx + \int_\mu^{\infty} (x-\mu)^2 f(x)\, dx,
\]
we must have
\[
\int_\mu^{\infty} (x-\mu)^2 f(x)\, dx \ge \frac{\sigma^2}{2}.
\]
Using this and (\ref{right}), we have (regardless of the magnitude of $l$),
\begin{equation}\label{right2}
\int_\mu^{\infty} (x-\mu)^2 f(x) \ge \frac{\sigma^2}{2^{10}}.
\end{equation}

Now we consider intervals to the right of $\mu$. Let $J_0 =
(\mu,x_0]$ where $x_0$ is the smallest point to the right of $\mu$
for which $f(x_0) \le 1/\sigma$ ($J_0$ could be empty). Let $J_i$,
for $i=1,2,\ldots,m=3\log (M/\sigma)+14$ be $[x_{i-1},x_i]$ where
$x_i$ is the smallest point for which $f(x_i) \le 1/(\sigma 2^i)$.
For any $t \ge t' \ge \mu$, $f(t') \ge f(t)F(t')/F(t) \ge f(t)F(\mu)
\ge f(t)/4$. Therefore, the function $f$ is approximately constant
in any interval $J_i$ for $i \ge 1$. If $x_0 > \mu+\sigma/64$, then
the interval $[\mu,\mu+\sigma/64]$ satisfies the desired property
(as $f(x) \geq f(x_0)$ for $x$ in this interval, we can take $\alpha
= f(x_0) \sigma/ 64 = 1/64$). Otherwise,
\[
\int_{J_0} (x-\mu)^2f(x)\, dx \le \sigma^2/2^{12}.
\]
Also,
\[
\int_{x_m}^\infty (x-\mu)^2f(x)\, dx \le  4 M^3 f(x_m) \leq
\sigma^2/2^{12}.
\]
Therefore, from (\ref{right2}), for some $i^* \geq
1$ we have
\[
\int_{J_{i^*}} (x-\mu)^2 f(x)\,dx \ge \frac{\sigma^2}{2^{12}m} .
\]
The interval $[\mu, x_{i^*}]$ then completes the proof: For this
interval we can take $\alpha = f(x_{i^*}) (x_{i^*} - \mu)$, and we
have
\begin{align*}
\int_{J_{i^*}} (x-\mu)^2 f(x)\,dx
    &\leq 8 (x_{i^*} - \mu)^2 (x_{i^*} - x_{i^*-1}) f(x_{i^*}) \\
    &\leq 8 \alpha (x_{i^*} - \mu)^2 .
\end{align*}
\end{proof}

%\begin{lemma}[variance of a product of independent variables]
%Let $X_1, \dotsc, X_n$ be independent random variables. Then
%\[
%\frac{\var (\prod_i X_i)}{(\E \prod_i X_i)^2} \geq \sum_i \frac{\var
%X_i}{(\E X_i)^2}.
%\]
%\end{lemma}
%\begin{proof}
%\begin{align*}
%\var (\prod_i X_i) &= \prod_i \E X_i^2 - (\prod_i \E X_i)^2 \\
%    &= \prod_i [\var X_i + (\E X_i)^2] - (\prod_i \E X_i)^2 \\
%    &= \prod_i (\E X_i)^2 [1+ \frac{\var X_i}{(\E X_i)^2}] - (\prod_i \E X_i)^2 \\
%    &= (\E \prod_i X_i)^2 [ \prod_i (1+ \frac{\var X_i}{(\E X_i)^2})-
%    1]\\
%    &\geq (\E \prod_i X_i)^2 \sum_i \frac{\var X_i}{(\E X_i)^2}.
%\end{align*}
%\end{proof}

%\begin{proposition}[variance of the product of the squared lengths, not used]
%Let $R_1, \dotsc, R_n \subseteq \RR^n$ be polytopes with at most
%$n^k$ facets, with $\vol R_i = 1$ and contained in the ball of
%radius $\sqrt{n}$. For all $i$ let $r_i \in R_i$ be random and
%independent. Then
%\[
%\var \prod_i \norms{r_i} \geq (\E \prod_i \norms{r_i})^2
%\frac{e^{-O(k)}}{\log n}
%\]
%\end{proposition}
%\begin{proof}
%By means of the previous lemma, Theorem \ref{thm:variancePolytope}
%and the fact that $\E \norms{r_i} \leq n$ we have
%\begin{align*}
%\var \prod_i \norms{r_i} &\geq (\E \prod_i \norms{r_i})^2 \sum_i
%\frac{\var \norms{r_i}}{(\E \norms{r_i})^2} \\
%    &\geq (\E \prod_i \norms{r_i})^2 \sum_i
%\frac{\var \norms{r_i}}{n^2} \\
%    &\geq (\E \prod_i \norms{r_i})^2 \frac{e^{-O(k)}}{\log n}.
%\end{align*}
%
%\end{proof}

\begin{proof}[Proof of Theorem \ref{PRODUCT}]
For this lower bound, we use the distribution $D'$ on matrices. Let
$R$ be an $n \times n$ random matrix having each entry uniformly and
independently in $[-1,1]$. On input $R$ from distribution $D'$
having rows $(R_1, \dotsc, R_n)$ and with probability at least $1/2$
over the inputs, we consider algorithms that output an approximation
to $f(R) = \prod_i \norm{R_i}$. The next claim for deterministic
algorithms, along with Yao's lemma, proves Theorem \ref{PRODUCT}.

\textbf{Claim:} Suppose that a deterministic algorithm makes at most
\[
    h:=\frac{\frac{n^2}{2} - 1}{\log_2 (2n+1)}
\]
queries on any input $R$ and outputs $V$. Then there exists a
constant $c>0$ such that the probability of the event
\[
\left(1-\frac{c}{\log n}\right)f(R) \le V \le \left(1+\frac{c}{\log
n}\right)f(R)
\]
is at most $1-O(1/n)$.

%\[
%\var(f(R)^2 \giventhat q) = (\E (f(R)^2 \giventhat q))^2
%\Omega(1/\log(n)).
%\]
%If $c$ is the output of the algorithm, this implies,
%\[
%\E (\frac{f(R)^2}{c^2} - 1)^2 \geq \Omega(\frac{1}{\log n})
%\]

To prove the claim, we consider a decision tree corresponding to a
deterministic algorithm. Let $P_l$ be the set of matrices associated
with a leaf $l$. By Lemma \ref{LEM:TREE}, we have that the set $P_l$
is a product set along rows, that is $P_l = \prod_i
\mathcal{R}_{i}$, where $\mathcal{R}_i \subseteq \RR^n$ is the set
of possible choices of the row $R_i$ consistent with $l$. The
conditional distribution of $R$ at a leaf $l$ consists of {\em
independent}, uniform choices of the rows from their corresponding
sets. Moreover, the sets $\mathcal{R}_i$ are polytopes with at most
$f=2n + 2h$ facets. Every query has at most $2n + 1$ different
answers, and every path has height at most $h$. Thus, $\card{L} \leq
(2n+1)^h =2^{\frac{n^2}{2} - 1}$. The total probability of the
leaves having probability at most $\alpha$ is at most $\alpha
\card{L}$. Thus, setting $\alpha = 1/(2\card{L})$, the leaves having
probability at least
\[
\frac{1}{2\card{L}} \geq \frac{1}{2^{n^2/2}}
\]
have total probability at least $1/2$. Because $\vol \cup_{l \in L}
P_l = 2^{n^2}$, we have that those leaves have volume at least
$2^{n^2/2}$. Further, since $P_l = \prod_i \mathcal{R}_i$, we have
that for such $P_l$ at least $n/2$ of the $\mathcal{R}_i$'s have
volume at least $1$. Theorem \ref{thm:variancePolytope} implies that
for those $\var \norms{R_i} \geq \Omega(n/\log n)$. Along with the
fact that $\norm{R_i} \leq \sqrt{n}$ and Lemma
\ref{lem:productVariance}, for a random matrix $R$ from such a
$P_l$, we get
\[
\frac{\var\bigl(f(R)^2\bigr)}{\bigl(\E (f(R)^2)\bigr)^2} \geq \sum_i
\frac{\var (\norms{R_i})}{\bigl(\E (\norms{R_i})\bigr)^2} =
\Omega\left(\frac{1}{\log n}\right).
\]
Thus, the variance of $f(R)$ is large. However, this does not
directly imply that $f(R)$ is dispersed since the support of $f(R)$
could be of exponential length and its distribution is not
logconcave.

%For the second conclusion, the fact that this happens with
%probability $1/2$ (i.e. for at least half of the leaves in
%probability) implies that after the algorithm runs we have also that
%$\frac{\var f(R)^2}{(\E f(R)^2)^2} \geq \Omega(1/\log n)$. This
%implies the second conclusion immediately by means of Lemma
%\ref{lem:varToExpectation}.
%

%\begin{lemma}\label{lem:varToExpectation}
%If $ \frac{\var X}{(\E X)^2} \geq \alpha \geq 0$ then for any $c \in
%\RR$ we have $\E (\frac{X}{c} - 1)^2 \geq \frac{\alpha}{1+\alpha}$.
%\end{lemma}
%\begin{proof}
%Equivalently, we want for $\beta = \frac{\alpha}{1+\alpha}$
%\[
%\E (X - c)^2 \geq  \beta c^2.
%\]
%The minimum of $\E (X - c)^2 - \beta c^2$ as a function of $c$ for
%fixed $\beta < 1$ is given by $2 c - 2 \E X = 2 \beta c$, that is,
%$c = \frac{\E X}{1-\beta}$. The hypothesis implies $\E X^2 \geq
%(1+\alpha) (\E X)^2$. Thus, for our choice of $\beta$,
%\begin{align*}
%\E (X - c)^2 - \beta c^2 &\geq \E (X - \frac{\E X}{1-\beta})^2 -
%\beta c^2 \\
%    &\geq (\E X)^2 [1+\alpha - \frac{1}{1-\beta}]\\
%    &= 0.
%\end{align*}
%\end{proof}
Let $X=\prod_{i=1}^n X_i$ where $X_i = \norms{R_i}$. To prove the
lower bound, we need to show that $\disp{X}{p}$ is large for $p$ at
least inverse polynomial in $n$. For $i$ such that
$\vol(\mathcal{R}_i) \ge 1$, we have $\var X_i = \Omega(n/\log n)$
by Theorem \ref{thm:variancePolytope}. As remarked earlier at least
$n/2$ sets satisfy the volume condition and we will henceforth focus
our attention on them. We also get
\begin{equation}\label{equ:expectationXi}
\E(X_i) \ge n/16
\end{equation}
from this. The distribution function of each $X_i$ is logconcave
(although not its density) and its support is contained in $[0,n]$.
So by Lemma \ref{UNIFORMPART}, we can decompose the density $f_i$ of
each $X_i$ as $f_i(x) = p_i g_i(x) + (1-p_i) g_i'(x)$. where $g_i$
is the uniform distribution over an interval $[a_i, b_i]$ of length
$L_i$ and
\[
p_iL_i^2 = \Omega\left(\frac{n}{\log^2 n}\right) \quad \mbox{ and } \quad p_i =
\Omega\left(\frac{1}{n\log^2 n}\right).
\]
We will assume that $p_iL_i^2 = cn/\log^2 n$ and $p_i =
\Omega(1/n^2)$. This can be achieved by noting that $L_i$ is
originally at most $n$ and truncating the interval suitably. Let
$X_i'$ be a random variable drawn uniformly from the interval
$[a_i,b_i]$. Let $Y_i = \log X'_i$, $I$ be a subset of
$\{1,2,\ldots,n\}$ and $Y_I=\sum_{i\in I} \log X'_i$. The density of
$Y_i$ is $h_i(t) = e^{t}/L_i$ for $\log a_i \le t \le \log b_i$ and
zero outside this range. Thus $Y_i$ has a logconcave density and so
does $Y_I$ (the sum of random variables with logconcave density also
has a logconcave density). Also, $\var(Y_I) = \sum_{i\in I}
\var(Y_i)$. To bound the variance of $Y_i$, we note that since $a_i
\ge \E(X_i) \ge n/16$ by Lemma \ref{UNIFORMPART} and Equation
\eqref{equ:expectationXi}, we have $b_i \le 16a_i$ and so $h_i(t)$
varies by a factor of at most $16$. Thus, we can decompose $h_i$
further into $h_i'$ and $h_i''$ where $h_i'$ is uniform over $[\log
a_i,\log b_i]$ and
\[
h_i(x) = \frac{1}{16}h_i'(x) + \frac{15}{16}h_i''(x).
\]
Let $Y_i'$ have density $h_i'$. Then
\[
\var(Y_i) \ge \frac{1}{16}\var(Y_i') = \frac{(\log b_i - \log
a_i)^2}{192}.
\]
Therefore
\[
\var(Y_I) \ge \frac{1}{192}\sum_{i \in I}(\log b_i - \log a_i)^2
\]
{}From this we get a bound on the dispersion of $Y_I$ using the
logconcavity of $Y_I$ and Lemma \ref{lem:varImpliesDisp}(b). The
bound depends on the set $I$ of indices that are chosen. This set is
itself a random variable defined by the decompositions of the
$X_i$'s. We have
\[
\e_I\bigl(\var(Y_I)\bigr) \ge \frac{1}{192}\sum_{i=1}^n p_i (\log
b_i - \log a_i)^2
 \ge \frac{1}{192}\sum_{i=1}^n p_i \frac{L_i^2}{(8a_i)^2} \ge \frac{c_1}{\log^2 n}
\]
On the other hand,
\begin{align*}
\var_I\bigl(\var(Y_I)\bigr) &\le \sum_{i=1}^n p_i(\log
b_i - \log a_i)^4\\
&\le \sum_{i=1}^n p_i
\frac{L_i^4}{a_i^4} \\
&\le \frac{16^4}{n^4}\sum_{i=1}^n \frac{p_i^2
L_i^4}{p_i}\\
&= \frac{16^4}{n^4}\frac{c^2 n^2}{\log^4 n}\sum_{i=1}^n
\frac{1}{p_i}.
\end{align*}
Suppose $p_i \ge c_2/n$ for all $i$. Then we get,
\[
\var_I\bigl(\var(Y_I)\bigr) \le \frac{c_2'}{\log^4 n}
\]
and for $c_2$ large enough, $\var_I\bigl(\var(Y_I)\bigr) \le
\bigl(\e_I \var(Y_I)\bigr)^2/4$. Hence, using Chebychev's
inequality, with probability at least $1/4$, $\var(Y_I) \ge
c_1/(4\log^2 n)$. By Lemma \ref{lem:varImpliesDisp}(b), with
probability at least $1/4$, we have $\disp{Y_I}{1/2} \ge
\frac{\sqrt{c_1}}{4\log n}$. This implies that for any $u$,
\[
\Pr\left(X \in \left[u, u\Bigl(1+\frac{\sqrt{c_1}}{4\log
n}\Bigr)\right]\right) \le \frac{7}{8}.
\]
Finally, if for some $i$, $p_i < c_2/n$, then for that $Y_i$, $L_i^2
=\Omega(n^2/\log^2 n)$ and using just that $i$, we get
$\disp{Y_i}{p_i/2} \ge \sqrt{L_i^2/a_i^2} = \Omega(1/\log^2 n)$ and
once again $X$ is dispersed as well (recall that $p_i = \Omega(1/n^2)$).
\end{proof}

\section{Variance of polytopes}\label{sec:VarPoly}

Let $X \in K$ be a random point in a convex body $K$.
Consider the parameter $\sigma_K$ of $K$ defined as
\[
\sigma_K^2 = \frac{n \var \norms{X}}{\bigl(\E \norms{X}\bigr)^2}.
\]
It has been conjectured that if $K$ is isotropic, then $\sigma_K^2
\leq c$ for some universal constant $c$ independent of $K$ and $n$
(the {\em variance hypothesis}). Together with the isotropic
constant conjecture, it implies Conjecture \ref{conj:variance}. Our
lower bound (Theorem \ref{thm:variancePolytope}) shows that the
conjecture is nearly tight for isotropic polytopes with at most
$\poly(n)$ facets and they might be the limiting case.
%\begin{theorem}[near tightness for variance hypothesis]
%Let $P \subseteq \RR^n$ be an isotropic polytope with at most $n^k$
%facets and let $X$ be a random point in $P$. Then
%\[
%    \var \norms{X} \geq c(k) \frac{n}{\log n}
%\]
%where $c(k)$ depends only on $k$ and not on $P$ or $n$.
%\end{theorem}

We now give the main ideas of the proof of
Theorem \ref{thm:variancePolytope}.
%In the rest we prove Theorem \ref{thm:variancePolytope}. An
%intuition that suggests that polytopes with few facets are different
%from the ball (say) is that most facets of such a polytope are large
%flat surfaces that are very different from the level sets of
%$\norms{\cdot}$, in other words, such a facet does not fit between 2
%concentric spheres of similar radii.
It is well-known that polytopes with few facets are quite different
from the ball. Our theorem is another manifestation of this
phenomenon: the width of an annulus that captures most of a polytope
is much larger than one that captures most of a ball. The idea of
the proof is the following: if $0 \in P$, then we bound the variance
in terms of the variance of the cone induced by each facet. This
gives us a constant plus the variance of the facet, which is a
lower-dimensional version of the original problem. This is the
recurrence in our Lemma \ref{lem:recurrence}. If $0 \notin P$ (which
can happen either at the beginning or during the recursion), we
would like to translate the polytope so that it contains the origin
without increasing $\var \norms{X}$ too much. This is possible if
certain technical conditions hold (case 3 of Lemma
\ref{lem:recurrence}). If not, the remaining situation can be
handled directly or reduced to the known cases by partitioning the
polytope. It is worth noting that the first case ($0 \in P$) is not
generic: translating a convex body that does not contain the origin
to a position where the body contains the origin may increase $\var
\norms{X}$ substantially. The next lemma states the basic recurrence
used in the proof.

\begin{lemma}[recurrence]\label{lem:recurrence}
Let $T(n,f,V)$ be the infimum of $\var\norms{X}$ among all polytopes
in $\RR^n$ with volume at least $V$, with at most $f$ facets and
contained in the ball of radius $R>0$. Then there exist constants
$c_1, c_2, c_3 >0$ such that
\begin{align*}
T(n,f,V) \geq \Bigl(1-\frac{c_1}{n}\Bigr) &T\biggl(n-1,f+2,
\frac{c_2}{nR^2}\Bigl(\frac{V}{Rf}\Bigr)^{1+\frac{2}{n-1}} \biggr) \\
&+ \frac{c_3}{R^{8/(n-1)}}
\left(\frac{V}{Rf}\right)^{\frac{4}{n-1}+\frac{8}{(n-1)^2}} .
\end{align*}
\end{lemma}
(Of course, $T$ depends on $R$, but we omit that dependence to
simplify the notation, given that, in contrast with the other
parameters, $R$ is the same for all appearances of $T$.)
\begin{proof}
Let $P$ be a polytope as in the statement (not necessarily minimal).
Let $U$ be the nearest point to the origin in $P$. We will use more
than one argument, depending on the case:

%\begin{enumerate}
\mycase{1} (origin) $0 \in P$.

For every facet $F$ of $P$, consider the cone $C_F$ obtained by
taking the convex hull of the facet and the origin. Consider the
affine hyperplane $H_F$ determined by $F$. Let $U$ be the nearest
point to the origin in $H_F$. Let $Y_F$ be a random point in $C_F$,
and decompose it into a random point $X_F+U$ in $F$ and a scaling
factor $t \in [0,1]$ with a density proportional to $t^{n-1}$. That
is, $Y_F = t(X_F+U)$. We will express $\var \norms{Y_F}$ as a
function of $\var \norms{X_F}$.

We have that $\norms{Y_F} = t^2 (\norms{U} + \norms{X_F})$. Then,
\begin{equation}\label{equ:varCone}
\begin{aligned}
\var \norms{Y_F}
%    &= \E (t^4 (\norm{U}^4+\norm{X_F}^4 + 2 \norms{U}
%    \norms{X_F})) - (\E (t^2 (\norms{U}+\norms{X_F})))^2 \\
%    &= \E t^4 (\norm{U}^4 + \E \norm{X_F}^4 + 2 \norms{U} \e
%    \norms{X_F}) - (\E t^2)^2 (\norms{U} + \E \norms{X_F})^2 \\
%    &= \E t^4 (\norm{U}^4 + \E \norm{X_F}^4 + 2 \norms{U} \e
%    \norms{X_F}) \\ &\quad - (\E t^2)^2 (\E \norms{X_F})^2 - (\E t^2)^2 \norm{U}^4
%    - 2 (\E t^2)^2 \norms{U} \E \norms{X_F}  \\
%%    &= (\E t^2)^2 \var \norms{X_F} + (\var t^2) \norms{U} +(\var t^2) \e
%%    \norm{X_F}^4 + 2 (\var t^2) \norms{U} \E \norms{X_F}
    = &(\E t^4) \var \norms{X_F} \\
    &+ (\var t^2) \bigl(\norm{U}^4 +(\e
    \norms{X_F})^2 + 2 \norms{U} \E \norms{X_F}\bigr)
\end{aligned}
\end{equation}
Now, for $k \geq 0$
\begin{align*}
\E t^k
%    = \frac{\int_0^1 t^k t^{n-1} dt}{\int_0^1 t^{n-1} dt}
    = \frac{n}{n+k}.
\end{align*}
and
\begin{align*}
\var t^2
%    &= \frac{n}{n+4} - (\frac{n}{n+2})^2 = \frac{n(n+2)^2 -
%    n^2(n+4)}{(n+4)(n+2)^2} \\
%    &= \frac{n^3 + 4n^2 + 4n - n^3 -4n^2}{(n+4)(n+2)^2} \\
    &=\frac{4n }{(n+4)(n+2)^2} \geq
    \frac{c_1}{n^2}
\end{align*}
for $c_1=1/2$ and $n \geq 3$. This in (\ref{equ:varCone}) gives
\begin{equation}\label{equ:varCone2}
\begin{aligned}
\var \norms{Y_F} &\geq \frac{n}{n+4} \var \norms{X_F} +
\frac{c_1}{n^2} \left(\norm{U}^4 + (\E \norms{X_F})^2
+ 2 \norms{U} \E \norms{X_F}\right) \\
&\geq \frac{n}{n+4} \var \norms{X_F} + \frac{c_1}{n^2} \bigl(\e
\norms{X_F}\bigr)^2.
%    \intertext{and, using Lemma \ref{lem:expectationVolume},}
\end{aligned}
\end{equation}
Now, by means of Lemma \ref{lem:expectationVolume}, we have that
\[
\E \norms{X_F} \geq c_2 V_{n-1}(F)^{2/(n-1)} (n - 1)
\]
and this in (\ref{equ:varCone2}) implies for some constant $c_3 > 0$
that
\[
\var \norms{Y_F} \geq \frac{n}{n+4} \var\norms{X_F} + c_3
V_{n-1}(F)^{4/(n-1)}.
\]
Using this for all cones induced by facets we get
\begin{equation}\label{equ:varCone3}
\begin{aligned}
\var\norms{X} &\geq \frac{1}{\vol P} \sum_{F \text{ facet}}
\vol{C_F} \var \norms{Y_F} \\
&\geq \frac{1}{\vol P} \sum_{F \text{ facet}} \vol{C_F}
\left(\frac{n}{n+4} \var \norms{X_F} + c_3
V_{n-1}(F)^{4/(n-1)}\right)
\end{aligned}
\end{equation}
Now we will argue that $\var \norms{X_F}$ is at least
$T(n-1,f,\frac{V}{R f})$ for most facets. Because the height of the
cones is at most $R$, we have that the volume of the cones
associated to facets having $V_{n-1}(F) \leq \vol P / \alpha$ is at
most
\[
    f \frac{1}{n} R \frac{\vol P}{\alpha}
\]
That is, the cones associated to facets having $V_{n-1}(F) > \vol P
/ \alpha$ are at least a $$1-\frac{Rf}{\alpha n}$$ fraction of $P$.
For $\alpha = R f$ we have that a $1-1/n$ fraction of $P$ is
composed of cones having facets with $V_{n-1}(F)
> \vol P /(Rf)$. Let $\mathcal{F}$ be the set of these facets.
The number of facets of any facet $F$ of $P$ is at most $f$, which
implies that for $F \in \mathcal{F}$ we have $$\var \norms{X_F} \geq
T(n-1,f,\frac{V}{R f}).$$ Then (\ref{equ:varCone3}) becomes
\begin{align*}
\var\norms{X} &\geq \frac{1}{\vol P} \sum_{F \in
\mathcal{F}} \vol{C_F} \left(\frac{n}{n+4}\var \norms{X_F} + c_3 V_{n-1}(F)^{4/(n-1)}\right) \\
    &\geq \frac{1}{\vol P} \sum_{F \in
\mathcal{F}} \vol{C_F} \left(\frac{n}{n+4} T\left(n-1,f,\frac{V}{R f}\right) + c_3 \left(\frac{V}{R f}\right)^{4/(n-1)}\right) \\
    &\geq \left(1-\frac{1}{n}\right) \left(\frac{n}{n+4}  T\left(n-1,f,\frac{V}{R f}\right) +
    c_3 \left(\frac{V}{R f}\right)^{4/(n-1)}\right) \\
    &\geq \left(1-\frac{c_5}{n}\right) T\left(n-1,f,\frac{V}{R f}\right) +
    c_4 \left(\frac{V}{R f}\right)^{4/(n-1)}
\end{align*}
for some constants $c_5, c_4>0$.

\mycase{2} (slicing) $$\var \e\bigl(\norm{X}^2 \giventhat
\inner{X}{U}\bigr) \geq \beta = \frac{c_4}{16} \left(\frac{V}{R
f}\right)^{4/(n-1)}.$$

In this case, using Lemma \ref{lem:SliceVar},
\begin{equation}\label{equ:slices}
\begin{aligned} \var \norms{X} &= \e
\var\bigl(\norms{X} \giventhat \inner{X}{U}\bigr) + \var
\e\bigl(\norm{X}^2 \giventhat \inner{X}{U}\bigr) \\
    &\geq \E \var\bigl(\norms{X} \giventhat \inner{X}{U}\bigr) + \beta
\end{aligned}
\end{equation}
Call the set of points $X \in P$ with some prescribed value of
$\inner{X}{U}$ a slice. Now we will argue that the variance of a
slice is at least $T\bigl(n-1,f,\frac{V}{2nR}\bigr)$ for most
slices. Because the width of $P$ is at most $2R$, we have that the
volume of the slices $S$ having $V_{n-1}(S) \leq V/\alpha$ is at
most $2 R V/\alpha$. That is, the slices having $V_{n-1}(S) >
V/\alpha$ are at least a $1-2R/\alpha$ fraction of $P$. For $\alpha
= 2 n R$, we have that a $1-1/n$ fraction of $P$ are slices with
$V_{n-1}(S) > V/(2n R)$. Let $\mathcal{S}$ be the set of these
slices. The number of facets of a slice is at most $f$, which
implies that for $S \in \mathcal{S}$ we have $\var\bigl( \norms{X}
\giventhat X \in S \bigr) \geq T\bigl(n-1,f,\frac{V}{2nR}\bigr)$.
Then (\ref{equ:slices}) becomes
\begin{align*}
\var \norms{X} &\geq \left(1-\frac{1}{n}\right)
T\left(n-1,f,\frac{V}{2nR}\right) + \frac{c_4}{16} \left(\frac{V}{R
f}\right)^{4/(n-1)}.
\end{align*}

\mycase{3} (translation) $\var (\inner{X}{U}) \leq \beta$ and $\var
\e\bigl(\norm{X}^2 \giventhat \inner{X}{U}\bigr) < \beta$.

Let $X_0 = X-U$. We have,
\begin{equation}\label{equ:translation}
\begin{aligned}
\var \norms{X}
%    &= \var (\norms{X_0} + 2 \inner{X_0}{U} + \norms{U}) \\
%    &= \var \norms{X_0} + 4 \var \inner{X_0}{U} + 4
%    \cov{\inner{X_0}{U}}{\norms{X_0}} \\
    &= \var \norms{X_0} + 4 \var \inner{X}{U} + 4
    \cov{\inner{X}{U}}{\norms{X_0}}.
\end{aligned}
\end{equation}
Now, Cauchy-Schwartz inequality and the fact that $\cov{A}{B} =
%\e(AB) - \e(A) \e(B) = \E (A \e(B \giventhat A)) - \e(A) \e(\E (B
%\giventhat A)) =
\cov{A}{\E (B \giventhat A)}$ for random variables $A,B$, give
\begin{align*}
\cov{\inner{X}{U}}{\norms{X_0}} &= \cov{\inner{X}{U}}{\norms{X}- 2
\inner{X}{U} + \norms{U}} \\
    &= \cov{\inner{X}{U}}{\norms{X}} - 2 \var \inner{X}{U} \\
    &= \cov{\inner{X}{U}}{\E (\norms{X} \giventhat \inner{X}{U})} - 2 \var \inner{X}{U} \\
    &\geq -\sqrt{\var \inner{X}{U}} \sqrt{\var \E (\norms{X} \giventhat \inner{X}{U})} - 2 \var
    \inner{X}{U}.
\end{align*}
This in (\ref{equ:translation}) gives
\begin{align*}
\var \norms{X}
    &\geq \var \norms{X_0} - 4 \var \inner{X}{U} -4 \sqrt{\var \inner{X}{U}} \sqrt{\var \E \bigl(\norms{X} \giventhat
    \inner{X}{U}\bigr)} \\
    &\geq \var \norms{X_0} - 8 \beta.
\end{align*}
Now, $X_0$ is a random point in a translation of $P$ containing the
origin, and thus case 1 applies, giving
\begin{align*}
\var \norms{X}
    &\geq \left(1-\frac{c_5}{n}\right) T\left(n-1,f,\frac{V}{R f}\right) +
    \frac{c_4}{2} \left(\frac{V}{R f}\right)^{4/(n-1)}.
\end{align*}

\mycase{4} (partition) otherwise:

We want to control $\var \inner{X}{U}$ to be able to apply the third
case. To this end, we will subdivide $P$ into parts so that one of
previous cases applies to each part. Let $P_1 = P$, let $U_i$ be the
nearest point to the origin in $P_i$ (or, if $P_i$ is empty, the
sequence stops), let $\hat U_i$ denote $U_i/\norm{U_i}$,
\[
Q_i = P_i \cap \left\{x \suchthat \norm{U_i} \leq \inner{\hat
U_i}{x} \leq \norm{U_i} + \sqrt{\beta}/R \right\},
\]
and $P_{i+1} = P_i \setminus Q_i$. Observe that $\norm{U_{i+1}} \geq
\norm{U_i} + \sqrt{\beta}/R$ and $\norm{U_i} \leq R$, this implies
that $i \leq R^2/\sqrt{\beta}$ and the
sequence is always finite. %Let $k$ be the length of the sequence.

For any $i$ and by definition of $Q_i$ we have $\var (\inner{X}{U_i}
\giventhat X \in Q_i) = \norms{U_i} \var (\inner{X}{\hat U_i}
\giventhat X \in Q_i) \leq \beta$.

The volume of the parts $Q_i$ having $\vol Q_i \leq V/\alpha$ is at
most $\frac{V R^2}{\alpha \sqrt{\beta}}$. That is, the parts having
$\vol Q_i
> V/\alpha$ are at least a $1-\frac{R^2}{\alpha \sqrt{\beta}}$
fraction of $P$. For $\alpha = n R^2 /\sqrt{\beta}$ we have that a
$1-1/n$ fraction of $P$ are parts with $\vol(Q_i) > V \sqrt{\beta}/
(n R^2)$. Let $\mathcal{Q}$ be the set of these parts. The number of
facets of a part is at most $f+2$. Thus, applying one of the three
previous cases to each part in $\mathcal{Q}$, and using that $f \geq
n$,
\begin{align*}
\var \norms{X}
    &\geq \frac{1}{\vol P} \sum_{Q \in \mathcal{Q}} \vol Q \var
(\norms{X} \giventhat X \in Q) \\
    &\geq \left(1-\frac{1}{n}\right) \left(\left(1-\frac{c_5}{n}\right) T\left(n-1,f+2,\frac{V \sqrt{\beta}}{nR^3 \max \{f, 2n\}}\right) +
\frac{c_4}{16} \left(\frac{V \sqrt{\beta}}{n R^3 f}\right)^{4/(n-1)}\right) \\
    &\geq \left(1-\frac{1}{n}\right) \left(\left(1-\frac{c_5}{n}\right) T\left(n-1,f+2,\frac{V \sqrt{\beta}}{2 f n R^3}\right) +
\frac{c_4}{16} \left(\frac{V \sqrt{\beta}}{n R^3
f}\right)^{4/(n-1)}\right).
\end{align*}

%\end{enumerate}

In any of these cases,
\begin{align}\label{equ:recurrenceCases}
\var \norms{X}
    &\geq \left(1-\frac{c_6}{n}\right) T\left(n-1,f+2,\frac{V}{2 Rf}\min\Bigl(1,\frac{\sqrt{\beta}}{nR^2}\Bigr)\right) +
c_7
\left(\frac{V}{Rf}\min\Bigl(1,\frac{\sqrt{\beta}}{nR^2}\Bigr)\right)^{4/(n-1)}.
\end{align}
Now, by assumption, $V \leq 2^n R^n$, and this implies by definition
that
\[
\frac{\sqrt{\beta}}{nR^2} \leq O\left(\frac{1}{n}\right).
\]
That is,
\[
\min\left(1,\frac{\sqrt{\beta}}{nR^2}\right) =
O\left(\frac{\sqrt{\beta}}{nR^2}\right)
\]
and the lemma follows, after replacing the value of $\beta$ in
Equation (\ref{equ:recurrenceCases}).
%, that
%there exists constants $c_6, c_8, c_9 > 0$ such that
%\begin{align*}
%\var \norms{X}
%    &\geq (1-\frac{c_6}{n}) T(n-1,f+2,c_8\frac{V}{ nR^3f} (\frac{V}{Rf})^{2/n}) +
%c_9 (\frac{V}{Rf})^{4/n} (\frac{(V/(Rf))^{1/n}}{R})^{8/n}.
%\end{align*}
\end{proof}

%\begin{theorem}[variance of a polytope]\label{thm:variancePolytope}
%Let $P \subseteq \RR^n$ be a polytope with at most $n^k$ facets,
%with $\vol P = 1$ and contained in the ball of radius $n^q$. Let $X
%\in P$ be a random point. Then
%\[
%\var\norms{X} \geq \frac{1}{e^{O(k+ q)}} \frac{n}{\log n}.
%\]
%\end{theorem}
%Note that there is no isotropy assumption.
\begin{proof}[Proof (of Theorem
\ref{thm:variancePolytope})] The inequality claimed in the theorem
is invariant under (uniform) scaling (which would change the volume
as well as the radius of the circumscribed sphere), and thus for the
proof we can assume that $\vol P = 1$, without loss of generality.
For $n\geq 13$, this implies that $R \geq 1$. We use the recurrence
lemma in a nested way $t=n/\log n$ times\footnote{To force $t$ to be
an integer would only add irrelevant complications that we omit.}.
The radius $R$ stays fixed, and the number of facets involved is at
most $f + 2t \leq 3f$. Each time, the volume is raised to the power
of at most $1+\frac{2}{n-t}$ and divided by at most
\[
u := c' nR^2 \bigl(R(f+2t)\bigr)^{1+\frac{2}{n-t}} > 1,
\]
for $c' = \max ( c_2^{-1}, 1 )$. That is, after $t$ times the volume
is at least (using the fact that $(1+\frac{2}{n-t})^t = O(1)$ and
denoting $v = 1+\frac{2}{n-t}$)
\begin{align*}
u^{-\sum_{i=0}^{t-1} v^i} \geq u^{-t v^t} =
\left(c' nR^2
\bigl(R(f+2t)\bigr)^{1+\frac{2}{n-t}}\right)^{-t(1+\frac{2}{n-t})^t}
    &\geq 1 / (3c'nR^3 f)^{O(t)}.
%    \\
%    &\geq 1 / [e^{(n\log 3 /\log n) + n + 3qn + kn} ]^{O(1)}.
\end{align*}
That means that from the recurrence inequality we get (we ignore the
expression in ``?'', as we will discard that term):
\begin{multline*}
T(n,f,1) \geq \left(1-\frac{c_1}{n}\right)^t T(n-t,f+2t,? ) + {} \\
{} + c_3 t \left(1-\frac{c_1}{n}\right)^{t-1}
\frac{1}{R^{8/(n-t-1)}} \left(\frac{1}{3Rf}\frac{1}{(3c'nR^3
f)^{O(t)}}\right)^{\frac{4}{n-1}+\frac{8}{(n-1)^2}} .
\end{multline*}
%We want to use the recurrence lemma in a nested way $t$ times. The
%radius $R$ stays fixed, and because $t \leq n$, the number of facets
%involved is at most $f + 2n = O(n^k)$. That means that the
%recurrence inequality becomes:
%\[
%T(n,f,V) \geq (1-\frac{c_6}{n}) T(n-1,f+2,\Omega(
%\frac{V^{1+\frac{2}{n}}}{nR^3 f} )) + c_9
%(\frac{V}{Rf})^{\frac{4}{n}+\frac{8}{n^2}} \frac{1}{R^{8/n}}.
%\]
%Every time it is used the number of facets increases by 2 and the
%volume is divided by $c_8 Rf = c_8 n^{kl}$. That is, after $t$
%applications the volume is at least $(1/n^{kl})^{n/\log n} \geq$
%%\begin{align*}
%%T(n,f,V) &\geq (1-\frac{c_6}{n-t})^t
%%T(n-t,f+2t,\frac{V}{[2\sqrt{n}(f+2t)]^t}) + t c_7
%%(1-\frac{c_6}{n-t})^t (\frac{V}{(f+2t)[2\sqrt{n}(f+2t)]^t})^{4/n}
%%\end{align*}
%\begin{align*}
%T(n,f,V) &\geq (1-\frac{c_6}{n-t})^t
%T(n-t,f+2t,\frac{V}{[2\sqrt{n}(f+2t)]^t}) + t c_9
%(1-\frac{c_6}{n-t})^t
%\end{align*}
We discard the first term and simplify to get,
\begin{align*}
T(n,f,1) &\geq  \frac{n}{\log n} \left(\frac{1}{R^3f
}\right)^{O(1/\log n)}
%    \\
%    &\geq  \frac{n}{\log n} e^{-O(q+k)}.
\end{align*}
Thus, for a polytope of arbitrary volume we get by means of a
scaling that there exists a universal constant $c > 0$ such that
\[
\var \norms{X} \geq (\vol P)^{4/n} \left(\frac{(\vol
P)^{3/n}}{R^3f}\right)^{c/\log n} \frac{n}{\log n}.
\]
The theorem follows.
\end{proof}

\section{Discussion}\label{DISCUSS}

The results for determinant/volume hold with the following stronger
oracle: we can specify any $k \times k$ submatrix $A'$ of $A$ and a
vector $x \in \R^k$ and ask whether $\inorm{A'x} \le 1$. In
particular, this allows us to query individual entries of the
matrix. More specifically, consider the oracle that takes indices
$i,j$ and $a \in \RR$ and returns whether $A_{ij} \leq a$. Using
this oracle, our proof (Lemma \ref{thm:determinantDispersion})
yields the following result: there is a constant $c>0$ such that any
randomized algorithm that approximates the determinant to within a
$(1+c)$ factor has complexity $\Omega(n^2)$. In the property testing
framework, this rules out sublinear (in the input size) methods for
estimating the determinant, even with randomized (adaptive) access
to arbitrary entries of the input matrix.

A posteriori, the way the volume lower bound is proved resembles an
idea used in communication complexity: discrepancy lower bounds. In
that idea, one gives an upper bound to the size of ``almost
monochromatic rectangles'', which implies a lower bound on the
number of rectangles and, thus, the communication complexity of the
given function. In our case, we give an upper bound to the measure
of product sets where the determinant does not change too much.
Moreover, our results imply a lower bound for the following
multi-party problem: There are $n$ players, player $i$ gets to know
only the $i$th row of a given $n \times n$ real matrix $A$, and they
want to approximate $\abs{\det{A}}$ up to a multiplicative constant.
Then in any protocol where each of them broadcasts bits, they must
broadcast $\Omega(n^2/\log n)$ bits, even for randomized protocols
succeeding with high probability and even if the matrix is
restricted to be far from singular as in Theorem
\ref{thm:determinantLowerBound}.

In our lower bounds for the product, the error bound is $1+c/\log n$,
where the logarithmic factor comes from the variance lemma. It is an
open problem as to whether this factor can be removed in the variance
lower bound.

For the volume problem itself, the best known algorithm has
complexity roughly $O(n^4)$ but the complexity of that algorithm is
conjectured to be $n^3$. It is conceivable that our lower bound for
membership oracle queries can be improved to $n^3$, although one
would have to use bodies other than parallelopipeds. Also, it is an
open problem to give a faster algorithm using a separation oracle.

Finally, we hope that the tools introduced here are useful for other
problems.

\medskip
\textbf{Acknowledgements.} We would like to thank Noga Alon, Laci
Lov{\'a}sz, Mike Saks and Avi Wigderson for helpful discussions.

\bibliographystyle{abbrv}    % bibliography using BibTex format
\bibliography{disp}

\end{document}

\end{document}